\documentclass[aps, pra, reprint, noshowpacs, twocolumn, superscriptaddress,nofootinbib,longbibliography
]{revtex4-2}
\usepackage{graphicx}
\usepackage{array}
\usepackage{url}
\usepackage{multirow}
\usepackage{soul}
\usepackage{xcolor}

\usepackage{mathtools,amssymb}
\usepackage{tikz}
\usepackage{tabularx}
\usepackage{enumitem}
\usepackage{censor}
\usepackage{svg}
\usepackage{import}
\usepackage[hang]{footmisc}
\newcommand{\mytop}{\rule{0 pt}{10pt}}
\newcommand{\mybottom}{\rule[-5pt]{0 pt}{5pt}}
\definecolor{crimson}{RGB}{220,20,60}
\newcommand{\hide}[1]{}

\newenvironment{iquote}
    {\vspace{0\baselineskip}%\itshape
    \list{}{\leftmargin=0.25in\rightmargin=0.25in}%
    \item\relax}
    {\endlist\vspace{0\baselineskip}}

\newcommand{\maps}{Maps}% 2
\newcommand{\blocks}{Blocks}% 4
\newcommand{\decay}{Decay}% 12
\newcommand{\mirrors}{Mirrors}% 13
\newcommand{\deltas}{Deltas}% 16
\newcommand{\efield}{E-Field}% 18

\begin{document}
\title{Investigating Unprompted and Prompted Diagrams Generated by Physics Majors During Problem Solving}

\author{Michael Vignal}
\affiliation{Department of Physics, University of Colorado, 390 UCB, Boulder, CO 80309}

\author{Bethany R. Wilcox}
\affiliation{Department of Physics, University of Colorado, 390 UCB, Boulder, CO 80309}

%--------------------------------- ABSTRACT ----------------------------
\begin{abstract}
Diagrams are ubiquitous in physics, especially in physics education and physics problem solving. Problem solvers may generate diagrams to orient to a scenario, to organize information, to directly extract an answer, or as a tool of communication. In this study, we interviewed 19 undergraduate and graduate physics majors, asking them to solve 18 multiple-choice physics problems---with no prompting regarding diagrams---and then six diagramming tasks of situations similar to six of the multiple-choice problems. By comparing spontaneously generated and prompted diagrams, we identify different diagramming elements and features used by physics majors acting towards different ends (\textit{i.e.,} in different epistemic frames). We found that different physical contexts impact how critical it is to draw an accurate diagram, and that the differences in diagramming between cohorts (\textit{e.g.}, between lower-division undergraduate and graduate students) seem to be smaller than the differences within a cohort. We also explore implications for teaching and research.
\end{abstract}

\maketitle

\newcommand{\correct}[1]{#1}
\newcommand{\one}{\hspace{-0.1em}}
\newcommand{\two}{\hspace{-0.2em}}

% %--------------------------------- INTRO ----------------------------

\section{Introduction and Background}
\label{sec:Introduction}

Diagrammatic representations of physical scenarios are ubiquitous in physics. Graphs and figures are used to communicate scientific information, such as in journal articles and textbooks. Graphs, sketches, and more specialized diagrams (\textit{e.g.}, free body diagrams, ray diagrams, etc.) can aid in physics problem solving. When problem solving, one may generate diagrams for a variety of reasons, including to orient to a situation or problem~\cite{scaife_external_1996, foster_implications_2002}; to aid in the problem-solving process~\cite{rosengrant_case_2006,kohl_expert_2007}; or as a tool of communication once a solution has been devised~\cite{heckler_consequences_2010,kirsh_thinking_2010}. Given the variety of diagrams and their uses in physics, it is not surprising that instruction around constructing diagrams is standard in physics education. Students are shown and expected to perform both general (\textit{e.g.}, graphs and sketches) and specific (\textit{e.g.}, free body diagrams, ray diagrams) techniques that can directly aid in solving physics problems~\cite{tairab_how_2004,meltzer_student_2005,nguyen_students_2009,savinainen_does_2013}.

Many physics education researchers have studied how students interact with diagrams. Most prevalent is research investigating how students interpret and use canonical and professional representations, for example, in the contexts of mechanics~\cite{dufresne_solving_1997,rosengrant_case_2006,heckler_consequences_2010,kohl_expert_2007,nguyen_facilitating_2010}, electricity and magnetism \cite{kohl_student_2005,mcpadden_network_2015}, quantum mechanics \cite{kohl_student_2005,gire_structural_2015}, thermodynamics \cite{bajracharya_students_2019}, and chemistry~\cite{hand_examining_2010}. Researchers have also used eye-tracking software to better understand how students refer to diagrams given with the problem prompts~\cite{susac_role_2019}. Much of this work specifically looks at the ways in which students interpret and coordinate between multiple representations~\cite{dufresne_solving_1997,rosengrant_case_2006,kohl_expert_2007,kohl_student_2005,nguyen_facilitating_2010,bajracharya_students_2019,mcpadden_network_2015,gire_structural_2015,hand_examining_2010,gire_multiple_2017}.

The majority of this work, however, has looked at how students use representations that have been provided to them, rather than representations the students generated themselves. Furthermore, few researchers have examined student-generated diagrams as artifacts that can help inform our understanding of student problem-solving~\cite{rosengrant_case_2006,kohl_expert_2007,heckler_consequences_2010,vignal_comparing_2020}, and most of this work has looked at diagrams collected as part of homework or exams, contexts which may influence how (and how often) students create diagrams. Some researchers have found that unprompted, accurate force diagrams may help students solve force problems~\cite{rosengrant_case_2006, heckler_consequences_2010}, but also that prompting for diagrams may interfere with student problem solving, to the detriment of student performance on those items~\cite{heckler_consequences_2010}. Additionally, researchers have found the characterization of student-generated diagrams difficult, especially while trying to avoid placing student work in a deficit framing~\cite{gupta_beyond_2011}. %There is no citation for this because it hasn't been published.

Expanded research into how and when students generate unprompted diagrams can inform a wide array of research and instructional practices: this work can inform research into student understanding of professional representations, including multiple representations; it can provide a foundation for more targeted investigations of student use of specific types of diagrams during problem solving; and it can help inform instructional practices regarding learning goals and assessment around student diagramming. It is in response to this need for research on spontaneously generated student diagrams that we situate our current study.

In a previous paper~\cite{vignal_comparing_2020}, we discuss the creation of a set of interviews designed to capture both unprompted and prompted student-generated diagrams. These diagrams were generated by 19 undergraduate and graduate physics majors during 1-on-1 problem solving interviews with author MV. A subset of the interview consisted of 6 multiple-choice problems (with no diagramming prompts) and paired diagramming tasks (with explicit diagramming prompts). By looking at both the unprompted and prompted diagrams generated by the same set of students on these six problem-task pairs, we attempted to answer the research questions:
\begin{iquote}
    \textbf{RQ1} How do spontaneously-generated student diagrams used in problem solving compare with similar, prompted student diagrams?
\end{iquote}
\noindent Given the variety of reasons for which a student might generate a diagram, a direct comparison between such diagrams can help us understand how diagrams generated without prompting, where students are generating diagrams for themselves, may differ from easier-to-capture prompted diagrams, where the student is generating the diagram for someone else. 

In this initial study, we found that students' unprompted diagrams were generally smaller, messier, and contained fewer details that their prompted diagrams \cite{vignal_comparing_2020}. Furthermore, we found instances in which unprompted diagrams helped students solve the problem (these were primarily problems that contained vectors), instances in which an unprompted diagram seemed unnecessary for the students to be able to correctly solve the problem, and instances in which unprompted diagrams did not seem to help students solve difficult problems. We concluded that diagramming can certainly be beneficial for students in some situations, but that there are many instances in which requiring students to diagram a problem may not be beneficial for them (at least in terms of getting the correct answer), which is consistent with and expands on previous findings that require students to draw free body diagrams doe not necessarily help them and may indeed impede students \cite{heckler_consequences_2010}.

We now expand on this work by looking at other aspects of student diagramming in both unprompted and unprompted situations, drawing data from all of the 18 problems and 6 diagramming tasks students worked though in the interviews described above. With this expanded data set, we aim to elaborate on \textbf{RQ1} and also explore the following additional research questions:
\begin{iquote}
    \textbf{RQ2} When and how do students use diagrams during problem-solving, such as to orient, organize information, or directly obtain an answer?
\end{iquote}
\begin{iquote}
    \textbf{RQ3} How does generation and use of diagrams differ between lower-division undergraduate, upper-division undergraduate, and graduate physics majors?
\end{iquote}

To answer \textbf{RQ2}, we look at what elements students include in their diagrams as well as when in the problem-solving process the diagram occurs. We identify instances when student-generated diagrams appear to be used primarily or entirely to orient the student to the scenario, as a tool that is manipulated to help solve the problem, or both.

To answer \textbf{RQ3}, we look at the various elements of our analysis (correct answers, accurate prompted diagrams, properties of unprompted diagrams, timing) across student cohorts ranging from 1st and 2nd year undergraduate physics majors to physics graduate students. Our findings with regards to this research question are informing ongoing research about the development of diagramming habits throughout a physicist's career. %I think we are seeing strong evidence, especially in the second round of interviews, that diagramming becomes a habit as people do physics more and more.

Development of our interview prompts is discussed in Sec.~\ref{sec:Methods}. Primarily qualitative analysis from our interviews are described in Sec.~\ref{sec:Results}, whereas quantitative analyses can be found in Sec.~\ref{quant}. Finally, Sec.~\ref{sec:Discussion} contains implications for teaching and research and also limitations of this study.

\section{Methods}
\label{sec:Methods}

To study and characterize diagrams generated by students, we conducted 19 problem-solving interviews in which physics majors worked on multiple-choice (MC) physics problems followed by explicit diagramming tasks. We interviewed 4 lower-division undergraduates (1st and 2nd year), 5 juniors (3rd year undergraduate), 5 seniors (4th year undergraduate), and 5 graduate students: this population of students was selected so we could explore diagramming throughout novice-to-expert development, and because graduate students are an understudied group in physics education research. Interview participants responded to email solicitations sent to physics majors at the University of Colorado Boulder and were financially compensated for their time.

The MC problems were primarily introductory-physics level content, and most could be solved entirely or in part with a diagram. One problem, with an electric charge distribution written in terms of delta functions, was not introductory level but was included as part of a potential expansion of previous findings of student difficulty with charge distributions described using delta functions~\cite{wilcox_student_2015}. Problem development began with the authors generating a list of content areas appropriate for physics majors who had taken introductory physics courses (Table~\ref{topics}): this process was guided by skimming through several introductory-texts \cite{knight_college_2015,giancoli_physics_2008}. Special attention was paid to topics with common or canonical diagrams. From this list, author MV developed 20 MC problems and 8 diagramming tasks that closely resembled 8 of the 20 MC problems. Although most of the MC problems were introductory level, they were intentionally designed to be challenging enough that we were confident students would be engaging in authentic problem solving.

\begin{table}[t]
    \caption{Topics covered in the interview. These topics were covered at the introductory level (with the exception of Delta functions, which was included to investigate student difficulties observed in previous work \cite{wilcox_student_2015}). Thermodynamics was abandoned due to the interview being too long and difficulty generating problems for which students were likely to spontaneously generate diagrams.}
    \centering
    \begin{tabular}{lll}\\\hline\hline\mytop
        Included    & Vector Addition           & Moment of Inertia \\
        Topics            & Kinematics                & Oscillations      \\
                    & Rotational Kinematics     & Mirrors           \\
                    & Forces                    & Lenses            \\
                    & Conservation of Energy    & Snell's Law       \\
                    & Conservation of Momentum  & Delta functions   \\
                    & Center of Mass            & Electric Potential\\\mybottom
                    & Torque                    & Electric Field    \\\hline
        Abandoned                               & Thermodynamics\\
        Topics\\\hline\hline
    \end{tabular}
    \label{topics}
\end{table}

To avoid inadvertently cuing students to draw diagrams during the interview, multiple-choice (MC) problem statements were text-only and did not ask for explanations or illustrations. Distractors were developed by intentionally making simple errors during the problem-solving process and listing the resulting answer as one of the MC options.

The diagramming tasks at the end of the interview asked students to carefully sketch/draw/graph and label a scenario resembling one of the MC problems, giving us problem-task pairs to compare unprompted and prompted diagrams of similar situations. These 28 items were piloted first by author BW, with follow-up pilots by an additional physics faculty member and a physics graduate student, both at the University of Colorado Boulder. The faculty member and graduate student were told that the purpose of the study was to investigate student problem solving, but we did not inform them that we were specifically interested in capturing and studying student-generated diagrams. Two thermodynamics problem-task pairs were abandoned after these pilot interviews over concerns of both interview length and the challenge of developing thermodynamic problems that were likely to illicit spontaneous diagrams from students without prompting. The remaining 18 MC problems and 6 diagramming tasks were tweaked following the pilot interviews, and the final wording of all problems and diagrams can be found in Tables \ref{ptpprompts} and \ref{pprompts}.

\begin{table*}
    \caption{The six problem (left) and task (right) prompts that comprised the problem-task pairs for our interview. The problems in this table are listed in the same order as they were presented to students with the exception of the \efield\ and \deltas\ problems, which were switched for students (though interspersed between these problems were the other 12 multiple-choice problems). The tasks were the final 6 items in the interview, presented in the same order as they were presented to students. Students were informed at the beginning of the interview that they could ask the interviewer for formulas and use a calculator during the interview. For the interviews, these problems were given numerical rather than descriptive names.}
    \label{ptpprompts}
\begin{tabularx}{\textwidth}{XcX}\hline \hline 
\multicolumn{3}{c}{\mytop\mybottom\textbf{Problem-Task Pairs}}\\\hline
\multicolumn{3}{c}{\mytop\textbf{\maps}}\\\mybottom
A car drives 2.5~miles South, 4~miles Southeast, 1~mile East, then 2~miles North. How far is the car from its starting point?
    \newline
    \textbf{a)} 4.1 miles \hfill
    \textbf{b)} \correct{5.1 miles} \hfill
    \textbf{c)} 6.3 miles\hfill
    \textbf{d)} 9.5 miles
&&
Carefully draw and label a `map' where a person  travels 200~m North, 200~m Southeast, 500~m South, then 400~m East.\\\hline
\multicolumn{3}{c}{\mytop\textbf{\blocks}}\\\mybottom
A 1 kg block sits on top of a 2 kg block, which sits on the floor. The coefficient of both static and kinetic friction is 0.4 between the two boxes and 0.3 between the bottom (2~kg) box and the floor. If a 100 N force is applied horizontally to the top (1 kg) box, will the bottom box slide along the~floor?
    \newline
    \textbf{a)} Yes \hfill
    \textbf{b)} \correct{No} \hfill
    \textbf{c)} Not enough information
&&
Carefully draw and label a free body diagram for a block sliding down a slope of angle $\theta$ with coefficient of kinetic friction~$\mu$.\\\hline
\multicolumn{3}{c}{\mytop\textbf{\decay}}\\\mybottom
A mass hanging from a spring is displaced 20 cm and oscillates up and down when released. Because of friction and air resistance, the amplitude of oscillation is halved every 10 seconds. What is the amplitude of oscillation after 15 seconds?
    \newline
    \textbf{a)} 4.6 cm \hfill
    \textbf{b)} 5.0 cm \hfill
    \textbf{c)} \correct{7.1 cm} \hfill
    \textbf{d)} None of the above
&&
Carefully sketch and label the graph of \mbox{$10\cos(x)\cdot2^{-\frac{x}{2\pi}}$} through at least 3 full periods.\\\hline
\multicolumn{3}{c}{\mytop\textbf{\mirrors}}\\\mybottom
Two flat, square mirrors are placed edge to edge with a 60$^\circ$ angle between their surfaces. Light comes in, bounces off of each mirror exactly once, and then leaves the system of mirrors. What is the angle between the incoming and outgoing~light?
    \newline
    \textbf{a)} 30$^\circ$\hfill
    \textbf{b)} 60$^\circ$\hfill
    \textbf{c)} \correct{120$^\circ$} \hfill
    \textbf{d)} Not enough information
&&
Carefully draw and label a ray-diagram for a ray of light that bounces off of two mirrors with an angle of 135$^\circ$ between them.\\\hline
\multicolumn{3}{c}{\mytop\textbf{\efield}}\\\mybottom
A charge of -$q$ sits at ($\ell$,0,0) and a second charge 2$q$ sits at (0,$\ell$,0). What is the electric field at (0,0,$\ell$)?
    \newline
    \textbf{a)}~$\dfrac{kq}{\ell^2} (\textrm{-}1,2,1)$ \hfill
    \textbf{b)}~$\dfrac{kq}{\ell^2} (1,\textrm{-}2,1)$ \hfill
    \textbf{c)}~$\dfrac{kq}{\ell^2} \dfrac{(\textrm{-}1,2,1)}{2^\frac{3}{2}}$ \hfill
    \textbf{d)}~\correct{$\dfrac{kq}{\ell^2} \dfrac{(1,\textrm{-}2,1)}{2^\frac{3}{2}}$} \vspace{1mm}
&&
Carefully draw and label 3 points: $A$ at ($\ell$,0,0), $B$ at \mbox{(0,-$\ell$,0)} and $C$ at (0,0,$\ell$). Then, if a charge -$q$ sits at $A$ and a charge $3q$ sits at $B$, sketch the electric field at the point $C$.\\\hline
\multicolumn{3}{c}{\mytop\textbf{\deltas}}\\\mybottom
Consider the 2-dimensional charge distribution:
        $$\ \quad \sigma(x,y)= A\ {\one}\delta(x{\one}-{\one}1)\delta(y{\one}+{\one}1) + B\ {\one}\delta(x{\one}+{\one}1)\delta(y{\one}-{\one}1)+C\ {\one}\delta(x{\one}+{\one}2),$$
        and assume $A$, $B$, and $C$ have the appropriate units to make all of the dimensions work out. How much total charge exists in the space defined below:
        $$-3 \leq x \leq 3 \qquad 0 \leq y \leq 3?$$
    \textbf{a)}~$A{\two}+{\two}B{\two}+{\two}C$ \hfill
    \textbf{b)}~$A{\two}+\hspace{-0.1em}B{\two}+{\two}3C$ \hfill
    \textbf{c)}~\correct{$B{\two}+{\two}3C$}\hfill
    \textbf{d)}~None of the above 
&&
Carefully draw and label the following charge distribution: \vspace{-0.4\baselineskip}
    $$\ \quad \sigma(x,y)=A\ {\one}\delta(x{\one}+{\one}2)\delta(y{\one}-{\one}1)+B\ {\one}\delta(x{\one}+{\one}1)\delta(y{\one}+{\one}2)+C\ {\one}\delta(x{\one}-{\one}1).\vspace{-0.6\baselineskip}$$\\\hline\hline
\end{tabularx}
\end{table*}

\begin{table*}[t]
    \caption{The 12 multiple-choice problems that were not paired with a diagramming task. These problems are listed in the same order (when read left to right, top to bottom) as they were presented to students, though they were interspersed between the problems from the problem-task pairs (table~\ref{ptpprompts}). For the interviews, these problems were given numerical rather than descriptive names.}
    \label{pprompts}
\begin{tabularx}{\textwidth}{XcX}\hline \hline 
\multicolumn{3}{c}{\mytop\mybottom\textbf{Unpaired Multiple-Choice Problem Prompts}}\\\hline 
\multicolumn{1}{c}{\mytop\textbf{Two Cars}}
&&
\multicolumn{1}{c}{\textbf{Projectile}}\\
Two cars drive from a house to a park. Both cars leave at the same time but take different routes. Car A travels 500 m at 10 m/s, then 3000 m at 30 m/s, and finally 1500 m at 20 m/s. Car B travels 2000 m at 10 m/s, then 1500 m at 20 m/s. Which car arrives at the park first?
    \newline
    \textbf{a)} \correct{Car A}\hfill
    \textbf{b)} Car B \hfill \
    \newline
    \textbf{c)} Car A and Car B arrive at the same time
&&
The position of a projectile is given by:
    $$y(x) = 5 \textrm{m} + 20 x - 2.5\ \textrm{m}^{-1} \cdot x^2.$$
    At what position $x$ does the projectile start moving in the negative $y$-direction?
    \newline
    \textbf{a)} 2 m\hfill
    \textbf{b)} \correct{4 m}\hfill
    \textbf{c)} 6 m \hfill
    \textbf{d)} 8 m
\\\hline 
\multicolumn{1}{c}{\mytop\textbf{Rolling Disk}}
&&
\multicolumn{1}{c}{\textbf{Stage Ramps}}\\ \mybottom
A solid disk of mass 1 kg and radius 0.1 m rolls along the ground at a speed of 10 m/s. The disk then rolls up a smooth hill. How high up the hill does the disk roll before stopping and rolling back down?
    \newline
    \textbf{a)} 5.0 m \hfill
    \textbf{b)} \correct{7.5 m} \hfill
    \textbf{c)} 10.0 m \hfill
    \textbf{d)} None of the above \hfill \
    \newline
    \textbf{e)} Not enough information
&&
An elevated stage has a steep ramp leading up to it on one side and a shallow ramp leading up to it on the other (the ramps are the same total height but different lengths). You want to slide a 100 kg box onto the stage: using which ramp requires you to expend the least energy to slide the box up and onto the stage?
    \newline
    \textbf{a)} The steep ramp \hfill
    \textbf{b)} The shallow ramp \hfill \
    \newline
    \textbf{c)} The amount of energy required is the same for both ramps
    \newline
    \textbf{d)} \correct{Not enough information}
\\\hline 
\multicolumn{1}{c}{\mytop\textbf{Center of Mass}}
&&
\multicolumn{1}{c}{\textbf{Collision}}\\ \mybottom
A uniform, flat tray (1 kg) centered at the origin holds three items: A book (2.5 kg) sitting at \mbox{(10~cm,~-10~cm)}; a cup (0.5 kg) sitting at (-10~cm,~-10~cm); and a plate (1 kg) sitting at (0 cm,~10 cm). What is the center of mass of the system (\textit{i.e.}, of the tray and the three objects)?
    \newline
    \textbf{a)} (0 cm, 0 cm) \hfill
    \textbf{b)} (0 cm, -2.5 cm) \hfill
    \textbf{c)} \correct{(4 cm, -4 cm)}
    \newline
    \textbf{d)} (5 cm, -5 cm)
&&
A 500 g basketball (at rest) is hit by a 50 g tennis ball moving at 8 m/s. If the tennis ball bounces off of the basketball at 2 m/s (back in the direction it came from), what is the speed of the basketball after the collision?
    \newline
    \textbf{a)} 2.4 m/s \hfill
    \textbf{b)} \correct{1.0 m/s} \hfill
    \textbf{c)} 0.8 m/s \hfill
    \textbf{d)} 0.6 m/s
\\\hline 
\multicolumn{1}{c}{\mytop\textbf{Circular Track}}
&&
\multicolumn{1}{c}{\textbf{Torque}}\\ \mybottom
A car drives around a circular track ($r=20$ m). The car starts from rest and accelerates at 1 m/s$^2$. What is the magnitude of centripetal (radial) acceleration after 10 seconds?
    \newline
    \textbf{a)} $\frac 12$ m/s$^2$ \hfill
    \textbf{b)} 1 m/s$^2$ \hfill
    \textbf{c)} 2 m/s$^2$ \hfill
    \textbf{d)} \correct{5 m/s$^2$}
&&
A tree branch of uniform mass density ($L=2$ m and $m=4$ kg) lies horizontally across a creek with only it's ends touching the ground. A squirrel ($m=500$ g) uses this branch as a bridge to cross over the creek. When the squirrel is one-third of the way across the branch, what is the upward force of the ground on the \textbf{far} end of the branch?
    \newline
    \textbf{a)} 20.83 N \hfill
    \textbf{b)} \correct{21.66 N} \hfill
    \textbf{c)} 22.50 N \hfill
    \textbf{d)} 23.33 N
\\\hline 
\multicolumn{1}{c}{\mytop\textbf{Moment of Interia}}
&&
\multicolumn{1}{c}{\textbf{Thin Lens}}\\ \mybottom
What is the moment of inertia for baseball bat of length $L$ (with a linear mass density $\lambda(x)=\lambda_0 + \alpha x$) rotating about $x=0$?
    \newline
    \textbf{a)} $\dfrac{\lambda_0 L^3}{3}$ \hfill
    \textbf{b)} $\dfrac{\lambda_0 L^3}{6}$ \hfill
    \textbf{c)} \correct{$\dfrac{\lambda_0 L^3}{3} + \frac{\alpha L^4}{4}$} \hfill
    \textbf{d)} $\dfrac{\lambda_0 L^3}{6} + \frac{\alpha L^4}{8}$
&&
A converging lens with a focal length of 1 m is placed 2 m from an object. What is the magnification $M$ of the image of the object?
    \newline
    \textbf{a)} 1 \hfill
    \textbf{b)} \correct{$-1$} \hfill
    \textbf{c)} $\dfrac 13$ \hfill
    \textbf{d)} $-\dfrac 13$
    \vspace{2pt}
\\\hline 
\multicolumn{1}{c}{\mytop\textbf{Snell's Law}}
&&
\multicolumn{1}{c}{\textbf{E-Potential}}\\ \mybottom
Light travelling in a vacuum strikes the window (n=1.3) of a spaceship at 10$^\circ$ from the normal. By what angle does the direction of the light change as it moves from the vacuum into the window?
    \newline
    \textbf{a)} 13$^\circ$ \hfill
    \textbf{b)} 7.7$^\circ$ \hfill
    \textbf{c)} \correct{2.3$^\circ$} \hfill
    \textbf{d)} None of the above
&&
What is the electric potential at a point 1 m from a +2 C charge, 2 m from a -3 C charge, and 3 m from a +1 C charge?
    \newline
    \textbf{a)} 0$k$ C/m \hfill
    \textbf{b)} \correct{$\frac {5}{6} k$ C/m} \hfill
    \textbf{c)} $\frac {5}{36} k$ C/m \hfill \
    \newline
    \textbf{d)} Not enough information
\\\hline\hline
\end{tabularx}\vspace{-8pt}
\end{table*}

As with the pilot interviews, students participating in the interviews were only told that we were investigating student problem solving; we gave them no indication we were interested in studying student diagrams. As a result, the diagrams that students generated while working through the MC problem were generated only if a student chose to draw the unprompted diagram as part of their problem-solving process. By asking students to generate prompted diagrams at the end of the interview, we were then able to compare student diagrams across epistemic frames~\cite{hammer_resources_2005, maries_impact_2016}. An epistemic frame is the perspective a learner has for an education setting or task. Working to learn and working to get the right answer are two common (and sometimes opposing) epistemic frames a student might adopt during problem solving. Another example of epistemic frames might be the idea that a task should be done independently or in collaboration with others. Studies of student diagrams that gather diagramming data from homework and exams, therefore, must be considered in the context of those assignments, where student diagrams are generally encouraged and often explicitly assessed.  

By telling our interview participants their goal was to get the correct answer, but by not asking them to show their work or draw diagrams, we aimed to place students in a problem-solving epistemic frame. This means that any diagram they generated would be generated because the student thought it would help them solve the problem, not because it would be expected or assessed.

In order to observe diagrams generated in an explicitly communicative epistemic frame, the final 6 items of the interview asked students to carefully generate and label a diagram: there was no actual problem to solve. By comparing these diagrams to students' unprompted diagrams, we could argue that differences in these diagrams suggest we were successful in capturing diagrams generated while students were in two distinct epistemic frames, and we could identify features of diagrams that appeared differentially in these different frames.

In tandem with the epistemic frames framework, we used the distributed cognition framework~\cite{salomon_practices_1997,foster_implications_2002,otero_conceptual_2002, kirsh_thinking_2010} to help us understand differences between students' unprompted and prompted diagrams. The distributed cognition framework, when applied to physics, holds that people will externalize (\textit{e.g.,} generate a diagram) for two main reasons: to reduce their cognitive load or to communicate information~\cite{salomon_practices_1997,foster_implications_2002}. It is reasonable to expect externalizations would differ when done for these two reasons. Furthermore, this framework helps us avoid a deficit framing of student diagrams: rather than view these diagrams as lacking when compared to expert diagrams, we can view the features that are present as signposts that help us understand students' cognitive processes.

\begin{figure}[t]
    \centering
    \includegraphics[width=\linewidth]{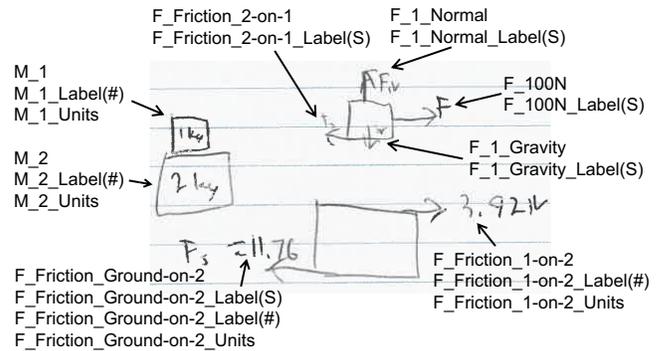}
    \caption{An illustration of the coding performed on all student diagrams. Actual coding was done in a spreadsheet and not as a mark-up of each diagram. We also timestamped each detail (excluding units) for the unprompted diagrams.}
    \label{coded_diagram}
\end{figure}

After conducting all 19 interviews, we reviewed the video recordings to code every mark and label students added to their unprompted and prompted diagrams. We did this by first generating a list of expected diagram elements (\textit{e.g.}, particular objects, arrows, labels, etc.), then iteratively coding these elements and identifying other marks and labels not yet accounted for in our coding~\cite{stemler_content_2015,cresswell_research_2014}. We revised our coding until it accounted for virtually every mark a student made as part of a diagram. As the marks coded required little qualitative interpretation (see Fig.~\ref{coded_diagram}), the authors examined a subset of diagrams to reach consensus after each round of coding, rather than seek to establish full inter-rater reliability (IRR). Coded elements were also timestamped based on the videos for the 18 MC problems. Timestamps were recorded using a custom computer script that allowed a researcher (author MV) to watch and control the video recordings manually while also capturing and recording timestamps into a spreadsheet, producing precise and accurate timestamps for more than 1700 diagram elements.

Using the timestamps of coded diagram elements, we also identified when during problem solving students generated diagrams and if students updated those diagrams. This added another lens through which to compare student-generated diagrams, investigating how diagram timing correlates with selecting the correct answer and how it differs across student cohort.

% From a resources perspective, we were interested to see how spontaneous diagrams could inform our understanding of which resources students activate in problem solving, though this is not discussed at length in this paper and will be part of a larger, upcoming study. We are also interested in what resources students activate in problem solving vs when the diagram is a tool of communication.
\section{Qualitative Results and Interpretations}
\label{sec:Results}

In this section, we will reiterate and elaborate on some of our previous findings~\cite{vignal_comparing_2020} regarding the 6 problem-task pairs (Sec.~\ref{problemtaskpairs}) and categorize the different ways that students used diagrams for each of the other 12 multiple-choice problems (Sec.~\ref{unpairedmcproblems}). In the following section~\ref{quant}, we look at the timing aspect of unprompted student-generated diagrams (\textit{i.e.}, when students drew) and other patterns that emerged in our analyses.  

\newcommand{\myarrow}{\rotatebox[origin=c]{180}{$\Lsh$}}

\begin{table}[t]\small
    \renewcommand{\arraystretch}{0.92}
    \vspace{-\baselineskip}
    \caption{Overall performance on problem-task pairs. Numbers are number of students. If either part of a problem-task pair was skipped due to time constraints, the pair is omitted from the table. Columns indicate correctness of student answers to problems. Sub-columns categorize \textit{unprompted} diagram content: more than given information~(G+); only given information~(G); no diagram~(ND); or blank~(B) if no work was shown. Rows describe accuracy of \textit{prompted} student diagrams: Accurate or containing small errors (e.g., a 400 m length drawn longer than a 500 m length), Inaccurate (e.g., forces missing or going in the wrong direction), or No Diagram if the student chose not to diagram the scenario.}
    \label{pairedperformance}
    \begin{tabularx}{\columnwidth}{lccccXccccXcccc}\hline\hline
    \multicolumn{3}{l}{\textbf{\maps}  (n=19)}\rule{0pt}{2ex}& \multicolumn{10}{c}{\maps~MC~Problem}    \\
    & \multicolumn{4}{c}{Correct} && \multicolumn{4}{c}{Incorrect}  && \multicolumn{4}{c}{No Answer} \\
    Paired~Task               & G+ & G  & ND & B  && G+ & G  & ND & B  && G+ & G  & ND & B \\
    \myarrow Accurate                 & 11 & 4  & -  & -  && 2  & -  & -  & -  && 1  & -  & -  & - \\
    \myarrow  Inaccurate              & -  & -  & -  & -  && -  & -  & -  & -  && -  & -  & -  & - \\
    \myarrow  No Diagram              & 1  & -  & -  & -  && -  & -  & -  & -  && -  & -  & -  & - \\\hline
    \multicolumn{3}{l}{\textbf{\blocks}  (n=19)}\rule{0pt}{2ex}& \multicolumn{10}{c}{\blocks~MC~Problem}    \\
    & \multicolumn{4}{c}{Correct} && \multicolumn{4}{c}{Incorrect}  && \multicolumn{4}{c}{No Answer} \\
    Paired~Task               & G+ & G  & ND & B  && G+ & G  & ND & B  && G+ & G  & ND & B \\
    \myarrow  Accurate                & 9  & 4  & -  & -  && 1  & 1  & -  & -  && 1  & 1  & -  & - \\
    \myarrow  Inaccurate              & -  & -  & -  & -  && -  & 1  & -  & -  && -  & 1  & -  & - \\
    \myarrow  No Diagram              & -  & -  & -  & -  && -  & -  & -  & -  && -  & -  & -  & - \\\hline
    \multicolumn{3}{l}{\textbf{\decay} (n=17)}\rule{0pt}{2ex}& \multicolumn{10}{c}{\decay~MC~Problem}    \\
    & \multicolumn{4}{c}{Correct} && \multicolumn{4}{c}{Incorrect}  && \multicolumn{4}{c}{No Answer} \\
    Paired~Task               & G+ & G  & ND & B  && G+ & G  & ND & B  && G+ & G  & ND & B \\
    \myarrow  Accurate                & 1  & 4  & 2  & 3  && -  & 1  & 1  & -  && -  & -  & -  & 1 \\
    \myarrow  Inaccurate              & 1  & -  & 1  & -  && -  & -  & -  & -  && -  & -  & -  & - \\
    \myarrow  No Diagram              & -  & 1  & -  & -  && -  & -  & -  & 1  && -  & -  & -  & - \\\hline
    \multicolumn{3}{l}{\textbf{\mirrors} (n=19)}\rule{0pt}{2ex}& \multicolumn{10}{c}{\mirrors~MC~Problem}    \\
    & \multicolumn{4}{c}{Correct} && \multicolumn{4}{c}{Incorrect}  && \multicolumn{4}{c}{No Answer} \\
    Paired~Task               & G+ & G  & ND & B  && G+ & G  & ND & B  && G+ & G  & ND & B \\
    \myarrow  Accurate                & 4  & 1  & -  & -  && 3  & 3  & -  & -  && 1  & -  & -  & - \\
    \myarrow  Inaccurate              & -  & 1  & -  & -  && 1  & -  & -  & -  && 1  & -  & -  & - \\
    \myarrow  No Diagram              & 2  & -  & -  & -  && -  & 1  & -  & -  && 1  & -  & -  & - \\\hline
    \multicolumn{3}{l}{\textbf{\efield} (n=15)}\rule{0pt}{2ex}& \multicolumn{10}{c}{\efield~MC~Problem}    \\
    & \multicolumn{4}{c}{Correct} && \multicolumn{4}{c}{Incorrect}  && \multicolumn{4}{c}{No Answer} \\
    Paired~Task               & G+ & G  & ND & B  && G+ & G  & ND & B  && G+ & G  & ND & B \\
    \myarrow  Accurate                & 4  & 1  & -  & -  && 3  & 1  & 1  & -  && -  & -  & 1  & - \\
    \myarrow  Inaccurate              & 1  & -  & -  & -  && -  & -  & -  & -  && -  & -  & -  & - \\
    \myarrow  No Diagram              & -  & -  & -  & -  && -  & 1  & -  & -  && 2  & -  & -  & - \\\hline
    \multicolumn{3}{l}{\textbf{\deltas} (n=18)}\rule{0pt}{2ex}& \multicolumn{10}{c}{\deltas~MC~Problem}    \\
    & \multicolumn{4}{c}{Correct} && \multicolumn{4}{c}{Incorrect}  && \multicolumn{4}{c}{No Answer} \\
    Paired~Task               & G+ & G  & ND & B  && G+ & G  & ND & B  && G+ & G  & ND & B \\
    \myarrow  Accurate                & -  & 2  & -  & -  && -  & 1  & -  & -  && -  & -  & -  & - \\
    \myarrow  Inaccurate              & -  & -  & -  & -  && -  & 2  & 2  & 1  && -  & -  & -  & - \\
    \myarrow  No Diagram              & -  & -  & 1  & -  && -  & 2  & 2  & 1  && -  & 1  & 1  & 2 \\\hline\hline
    \end{tabularx}
\end{table}

\subsection{Problem-Task Pairs}
\label{problemtaskpairs}

Table~\ref{pairedperformance} shows student performance on problem-task pairs across three axes: answer correctness, unprompted diagram detail, and prompted diagram accuracy. For unprompted diagrams, we identify diagrams depicting only information given in the prompt~(G), diagrams depicting additional information, such as calculated values or simplifications~(G+), when no diagram was drawn~(ND), and when the work area was left blank~(B). Distinguishing between~(G+) and~(G) helped us determine if a diagram was used for organizing information, discussed more in Sec.~\ref{quant}. We now discuss student answers and diagrams for each problem-task pair.

% What was typically drawn
% Answers
% Comparison between unprompted and prompted

\subsubsection{\maps}
The \maps\ problem-task pair (described in Table \ref{ptpprompts}) required students to consider the addition of four spatial vectors. As measured by the high number of correct student answers (Table \ref{pairedperformance}), this problem was one of the easiest in the interview and the easiest of the 6 problems that were paired with a diagramming task. Every student drew an unprompted diagram for the multiple-choice problem. For the diagramming task, every student completed the task and only one student drew an inaccurate diagram. Only 3 students answered the problem incorrectly: 2 because of algebra mistakes and 1 who only drew 3 of the 4 segments of the path. No students attempted to answer the problem by drawing a diagram to scale and measuring the desired distance. 

\begin{figure}[t]
    \centering
        \includegraphics[width=0.468\linewidth]{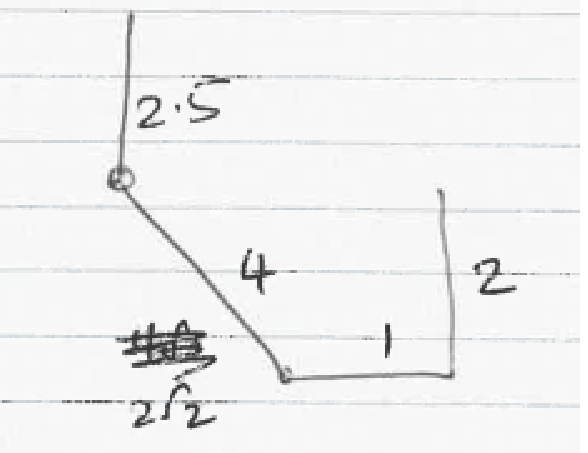}
        \includegraphics[width=0.452\linewidth]{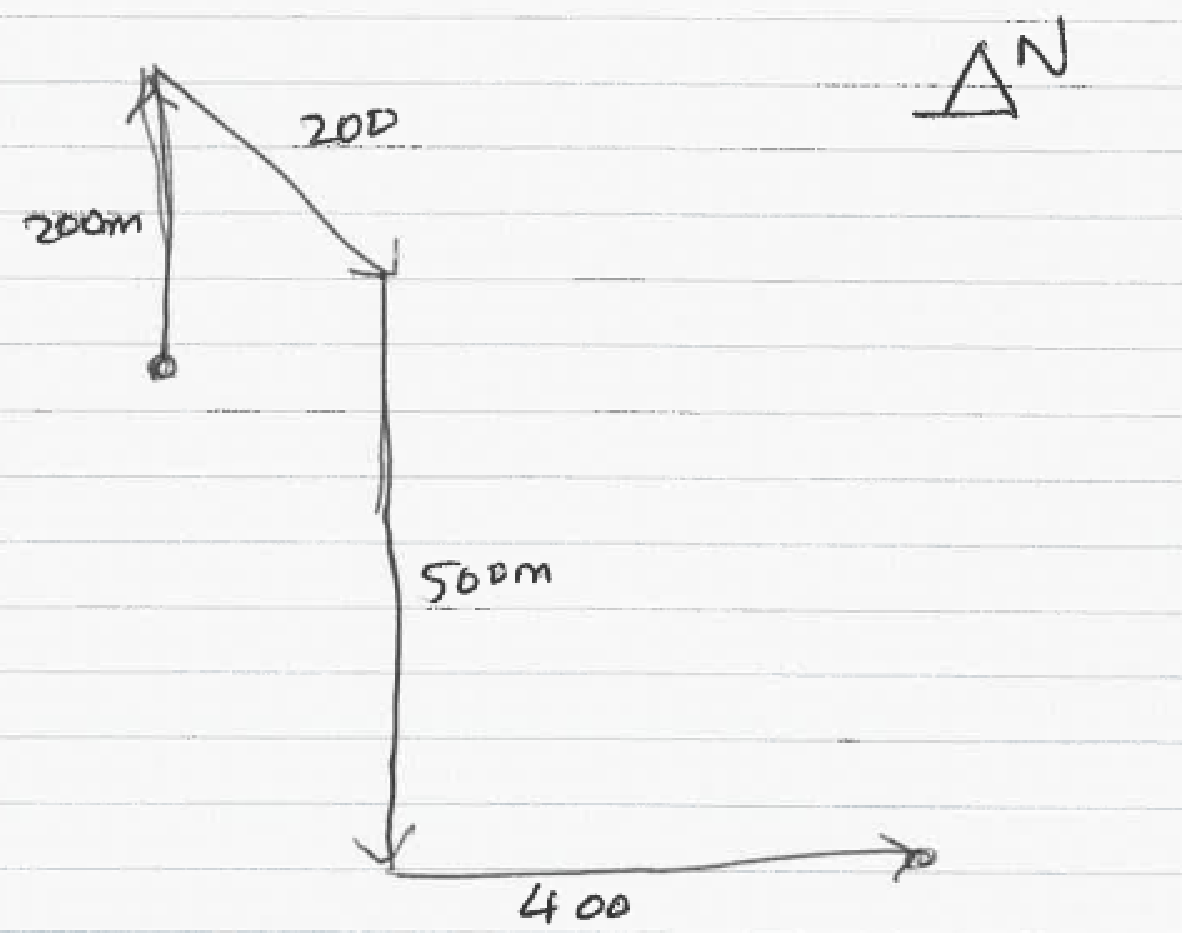}\\
        (a) \maps\ diagrams for one student.\\
        \includegraphics[width=0.402\linewidth]{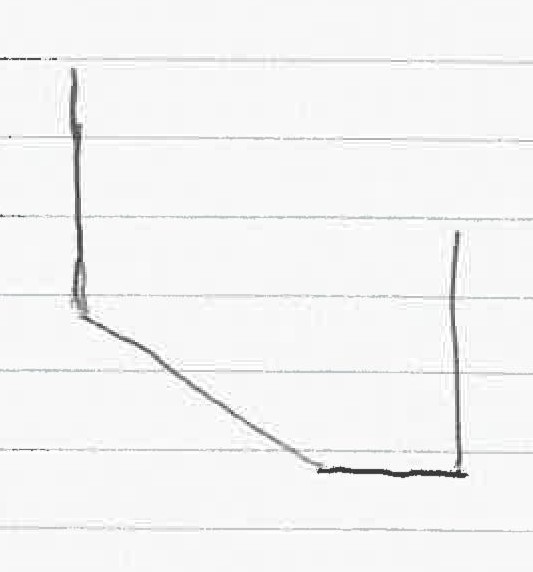}
        \includegraphics[width=0.518\linewidth]{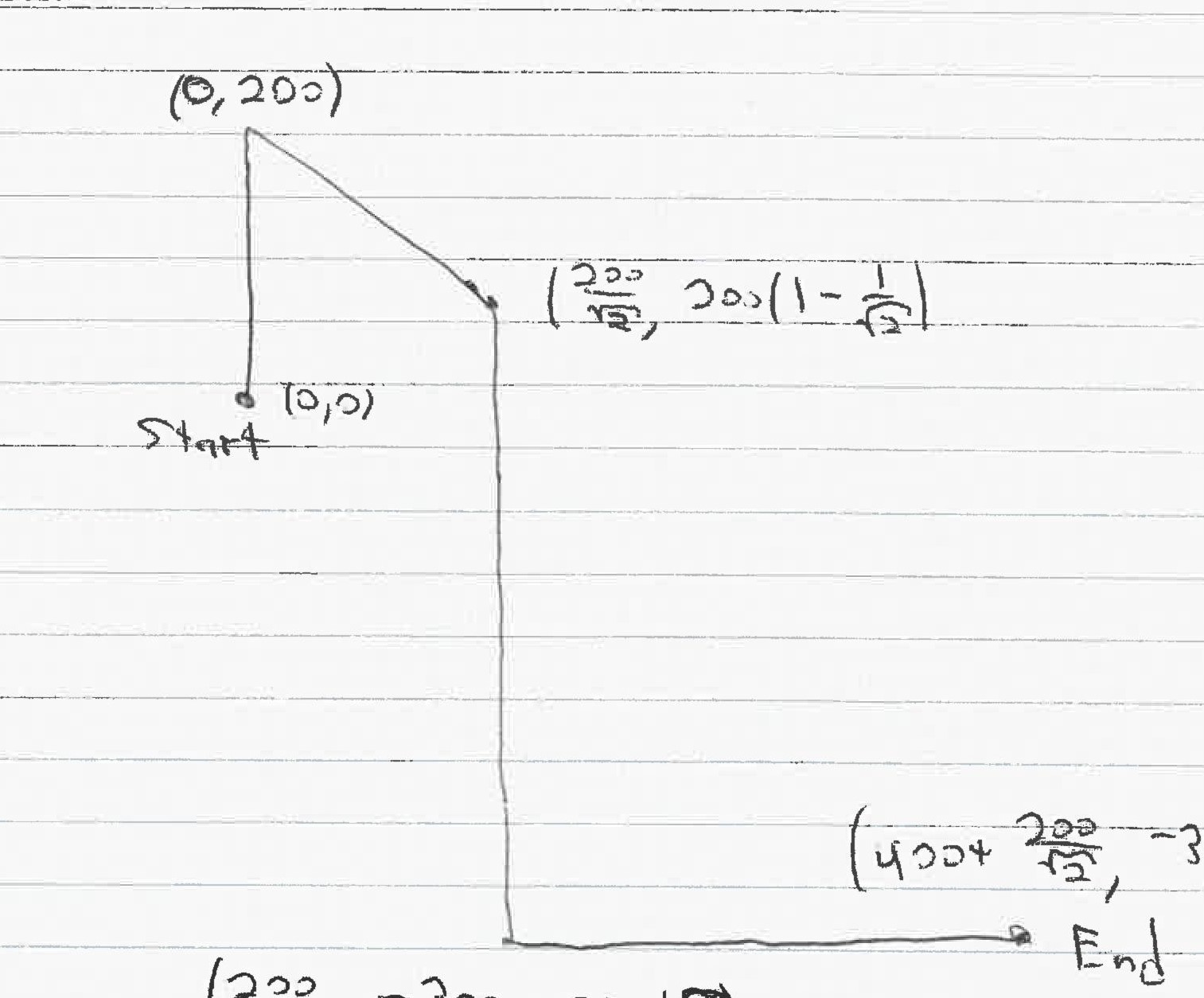}\\
        (b) \maps\ diagrams for another student.
        \caption{Examples of unprompted (left) and prompted (right) student diagrams for the maps problem-task pair, with each pair of images being from one student. Detail of unprompted diagrams (left) varied widely.}
        \label{mapdiagrams}
\end{figure}

As can be seen in Fig. \ref{mapdiagrams} -- and this holds for all 18 problems the students were asked to complete -- the level of detail in the prompted diagram could vary wildly. Of the 6 problems discussed in this paper, the \maps\ problem is the only one where every student clearly referred to their unprompted diagram when answering the question (specifically we observed all students retrieving numeric values from the diagram while setting up their algebraic solutions). So while the level of detail could vary, all of the students used the diagrams to answer the question, and almost every student (including the two whose diagrams are shown in Fig.~\ref{mapdiagrams}) answered the problem correctly. This could indicate that the drawn segments are all students needed to externalize to answer the question or that different students needed to externalize different amounts in order to succeed (and they only drew what was needed). In reality, we believe the truth is somewhere in between, that the segments and potentially other features needed to be externalized, but some of the features were likely not needed to solve the problem.

\subsubsection{\blocks}
The \blocks\ problem-task pair asked students to consider mechanical forces (including friction) acting on massive blocks. The problem is a canonical but challenging force problem that requires students to reason about and correctly calculate friction forces, and for this problem, every student drew an unprompted diagram, though 8 students only included information that was given to them in the problem statement (\textit{i.e.}, they did not add any force-pairs, calculated values, or other features not described in the problem statement). The presence of (and detail in) unprompted diagrams is not strongly correlated with the student selecting the correct answer, as can be seen in Table~\ref{pairedperformance} and discussed more in Sec. \ref{quant}, with the exception that 5 of 6 students who drew the (non-given) forces that were acting on the block in question (the bottom block) got the correct answer.

\begin{figure}[t]
    \centering
        \includegraphics[width=0.460\linewidth]{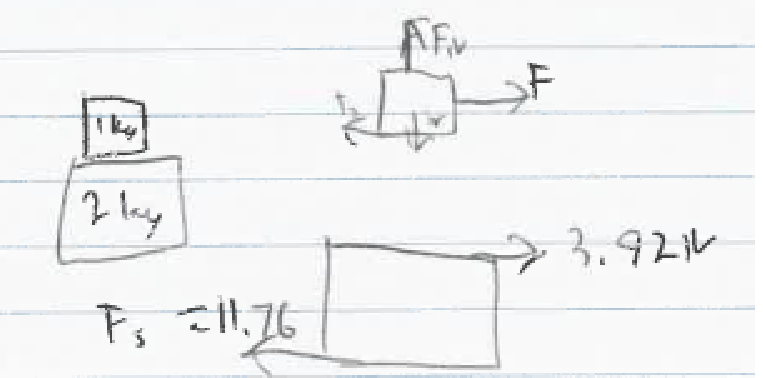}
        \includegraphics[width=0.460\linewidth]{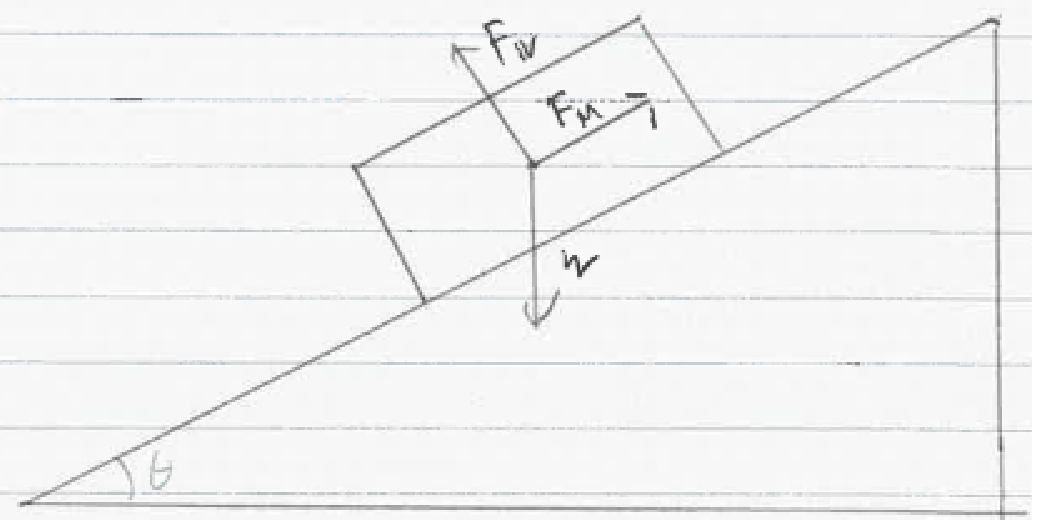}\\
        (a) \blocks\ diagrams for one student.\\
        \includegraphics[width=0.243\linewidth]{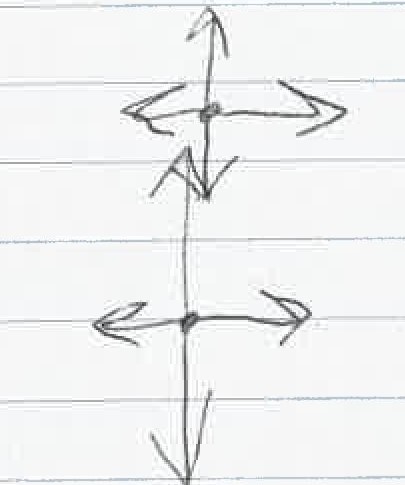}
        \includegraphics[width=0.677\linewidth]{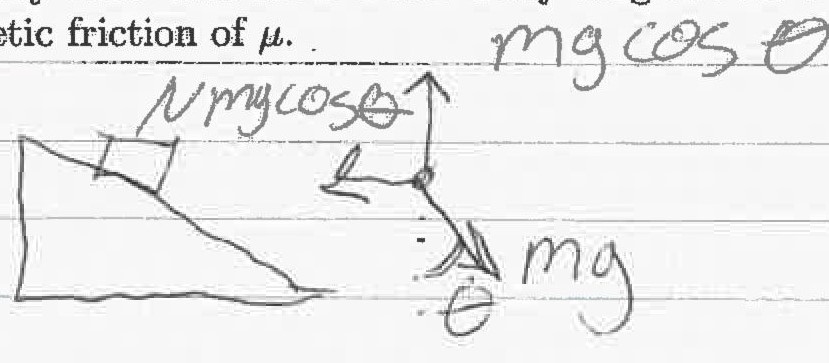}\\
        (b) \blocks\ diagrams for another student.\\
        \caption{Examples of unprompted (left) and prompted (right) student diagrams for the blocks problem-task pair, with each pair of images being from one student.}
        \label{blocksdiagrams}
\end{figure}

In the unprompted diagrams, 11 of 14 students who represented forces with arrows drew those arrows at the locations where the forces act (Fig.~\ref{blocksdiagrams}), rather than at the center of mass, and 5 students also did this during the prompted diagramming task. We note this is consistent with student examples shown in Heckler~\cite{heckler_consequences_2010}, though this feature is not discussed at length in that text.

As exemplified in Fig.~\ref{blocksdiagrams}, the amount of detail in unprompted diagrams varied greatly. For this problem, diagrams that include only information given in the problem statement do not seem to improve the chances a student will answer the question correctly. The only details that appear to increase the likelihood a student will select the correct answer are the non-given forces on the bottom block. Many students were able to draw accurate free-body diagrams for forces acting on blocks when they were prompted to do so but chose not to for the MC problem. This suggests that students do not automatically draw diagrams (or elements of diagrams) that could help them correctly solve problems, even if they are capable of drawing such diagrams.

% What was typically drawn
% Answers
% Comparison between unprompted and prompted

\subsubsection{\decay}

\begin{figure}[t]
    \centering
        \includegraphics[width=0.386\linewidth]{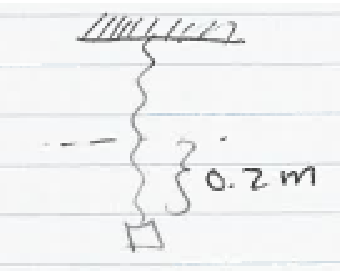}
        \includegraphics[width=0.534\linewidth]{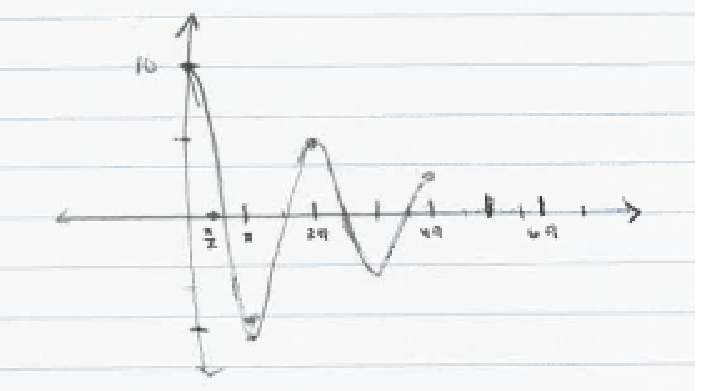}\\
        (a) \decay\ diagrams for one student.\\
        \includegraphics[width=0.462\linewidth]{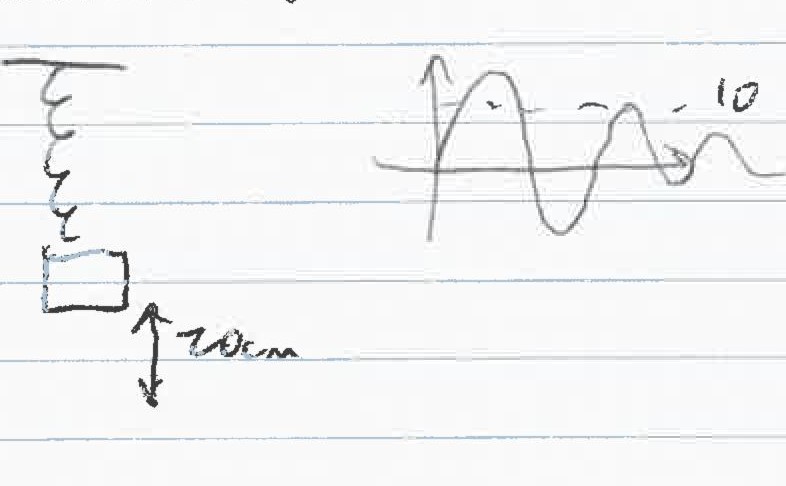}
        \includegraphics[width=0.458\linewidth]{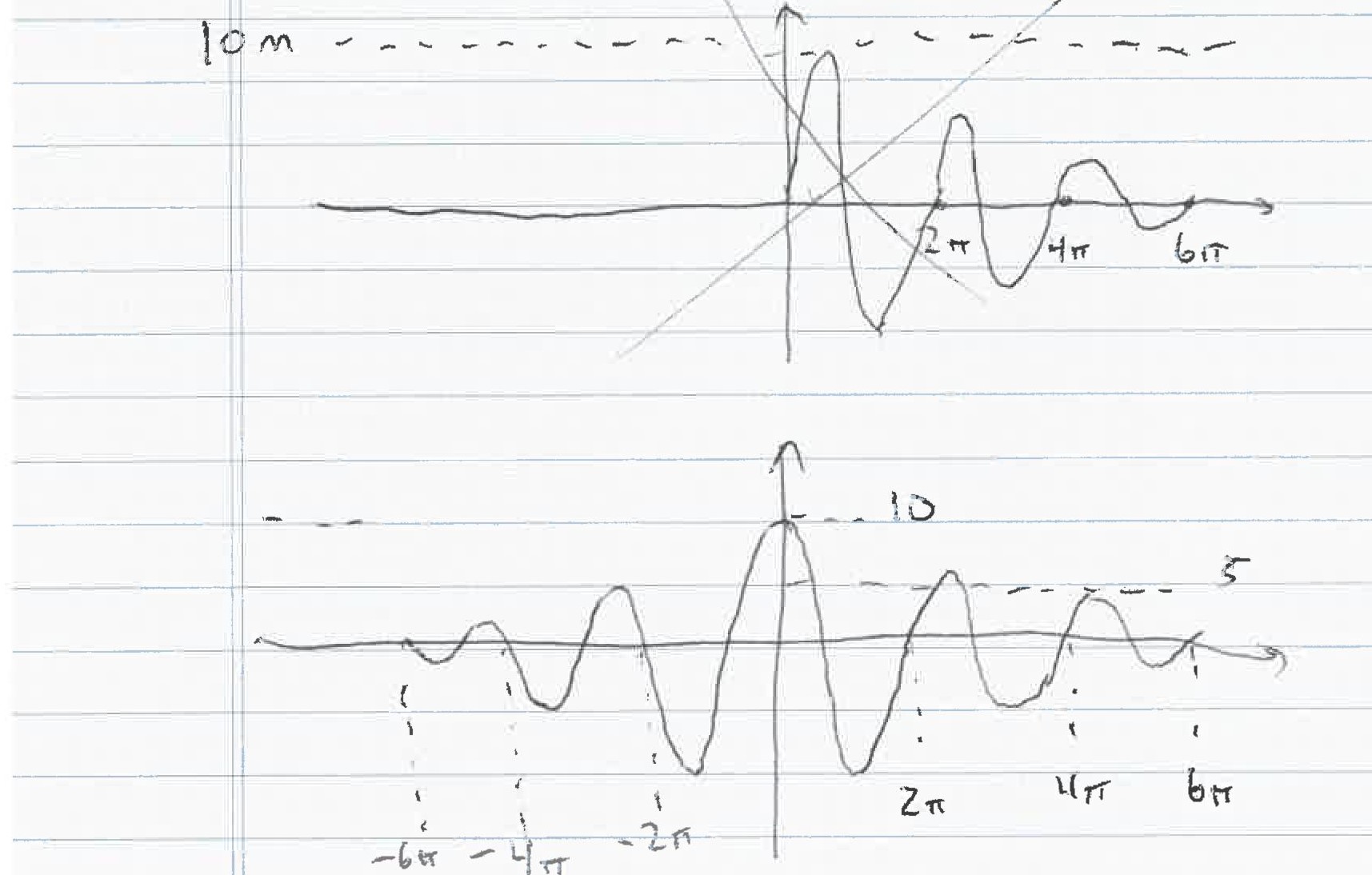}\\
        (b) \decay\ diagrams for another student.\\
        \caption{Examples of unprompted (left) and prompted (right) student diagrams for the decay problem-task pair, with each pair of images being from one student.}
        \label{decaydiagrams}
\end{figure}

The \decay\ scenarios asked students to consider the amplitude (over time) of a damped oscillator. While it is possible to plot the displacement of the oscillator over time (an exponentially decaying cosine curve) or depict the amplitude of the oscillator over time (a decaying exponential), only one student drew an unprompted graph depicting the oscillator (Fig.~\ref{decaydiagrams}(b) left) for the multiple choice problem. Eleven students, including the one who drew a graph, drew sketches of the oscillator (\textit{e.g.}, Fig.~\ref{decaydiagrams}(a) left and (b) left). Only two students included any features in their diagrams that were not stated in the problem statement.

This problem-task pair provides strong evidence of diagrams being used by students almost exclusively to orient them to the physical situation, as a single sketch of the oscillator does not capture the time evolution of this system and thus cannot directly help students answer the question. This is not to say that the sketches were not useful or that orienting a student to a problem is not a productive use of a diagram, only that these diagrams did not directly lead the student to a solution. As 13 of 17 students who answered the question selected the correct answer, and as students were largely successful in drawing a decaying trig function when prompted (\textit{e.g.}, Fig.~\ref{blocksdiagrams}(a) right and (b) right), we believe students productively chose tools other than diagramming (\textit{e.g.}, algebra, sensemaking) to solve this problem, and thus graphs of the oscillator were unnecessary.

Furthermore, while students' prompted diagrams of the decaying oscillator generally had the correct shape, many students struggled with these diagrams. As can be seen in Fig.~\ref{blocksdiagrams}(b), some students struggled with knowing whether decaying sine or cosine curves were appropriate or with knowing what the graph would look like for $t<0$. This suggests the possibility that, had we required students to generate diagrams as part of solving the MC, difficulties with these diagrams could have negatively impacted student's ability to answer the question correctly.

% What was typically drawn
% Answers
% Comparison between unprompted and prompted
\subsubsection{\mirrors}

\begin{figure}[t]
    \centering
        \includegraphics[width=0.376\linewidth]{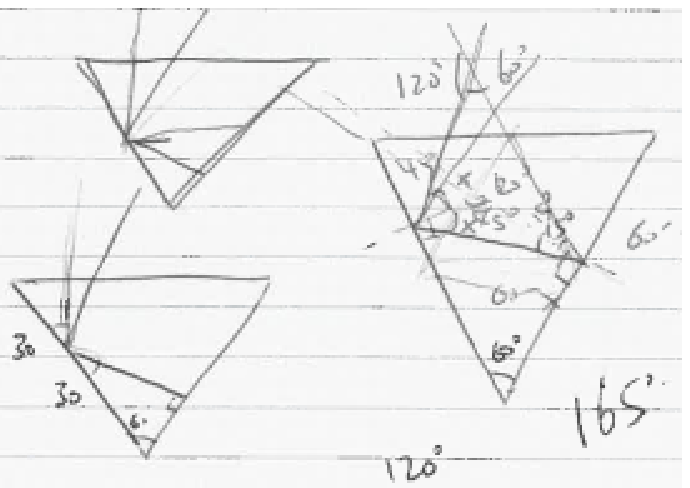}
        \includegraphics[width=0.544\linewidth]{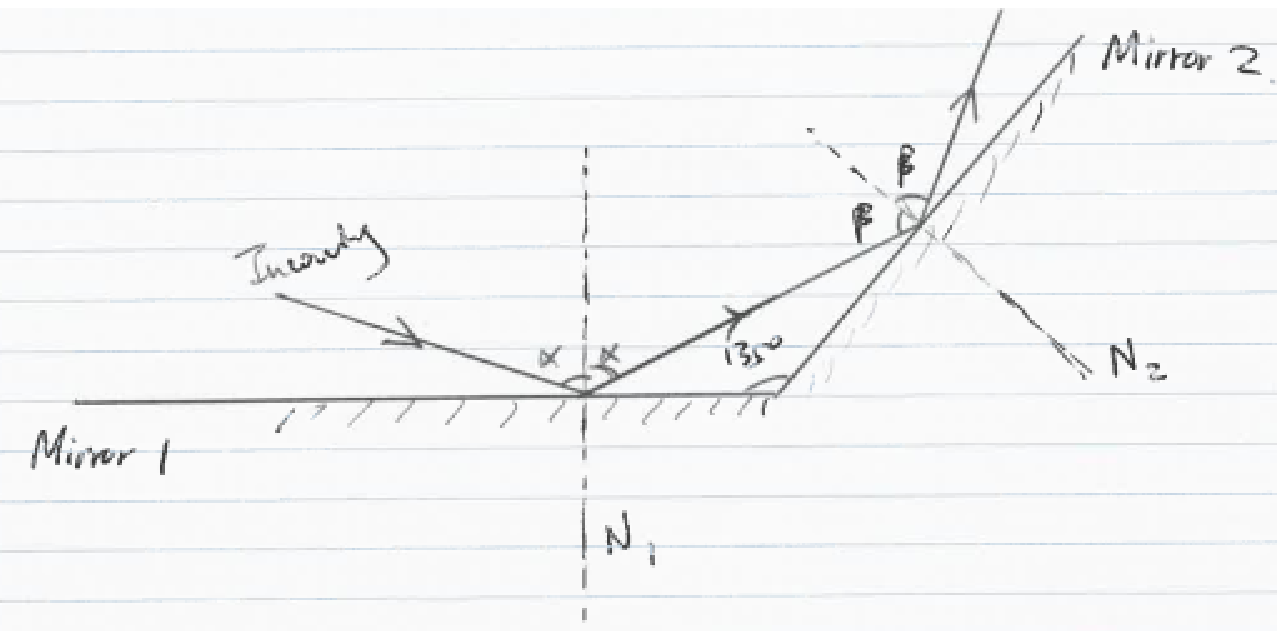}\\
        (a) \mirrors\ diagrams for one student.\\
        \includegraphics[width=0.380\linewidth]{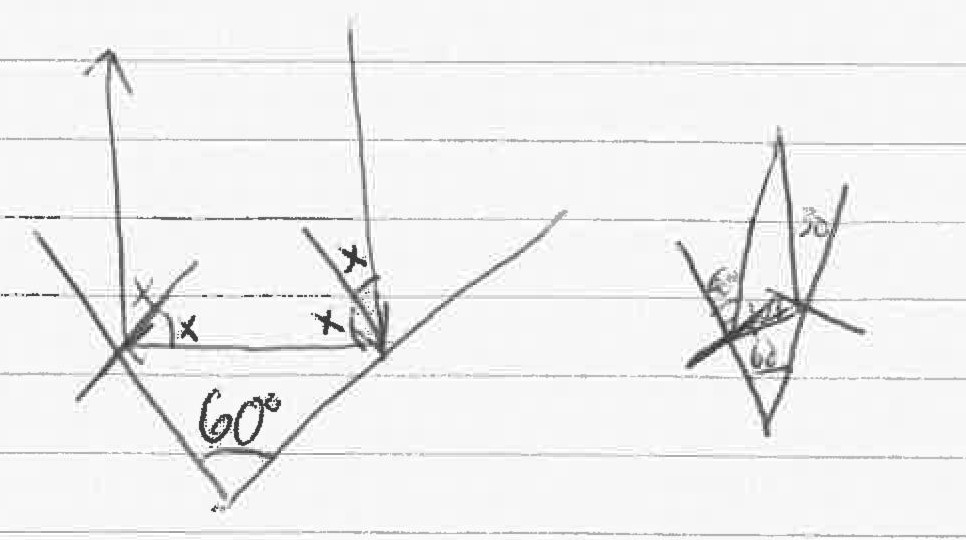}
        \includegraphics[width=0.540\linewidth]{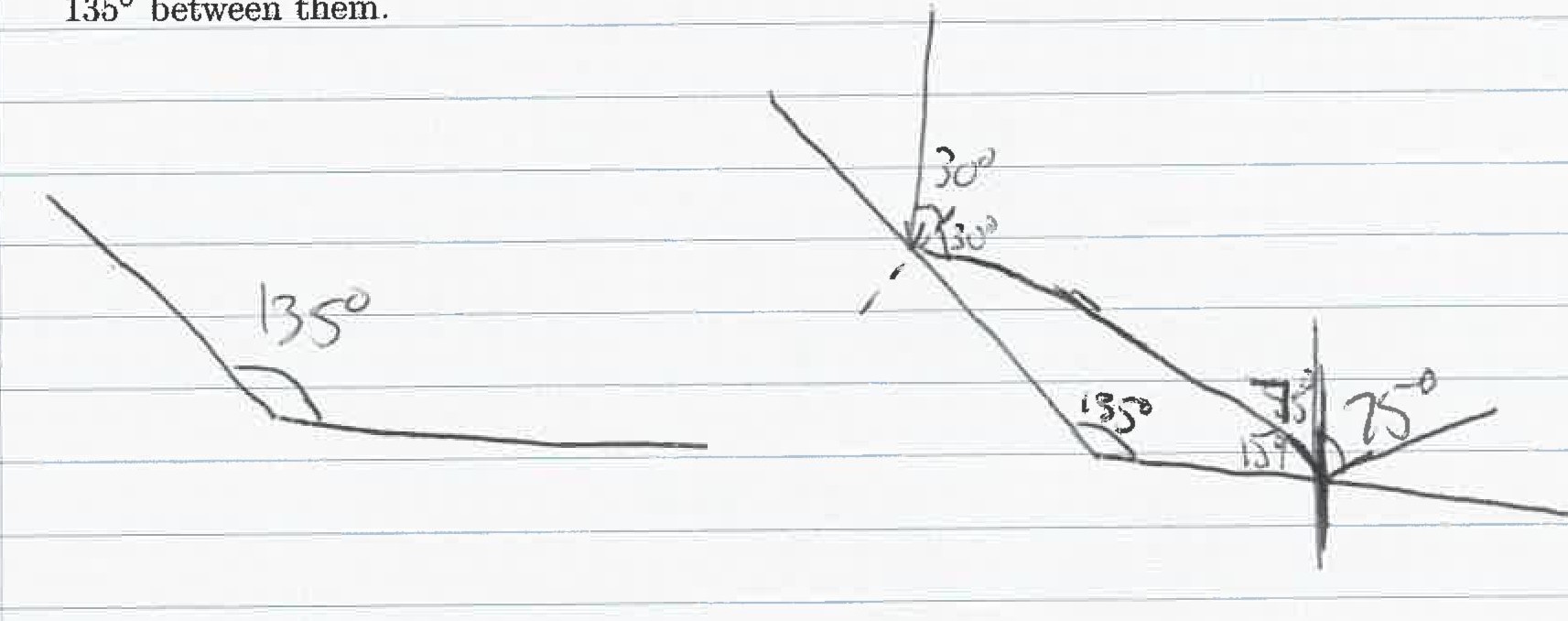}\\
        (b) \mirrors\ diagrams for another student.
        \caption{Examples of unprompted (left) and prompted (right) student diagrams for the mirrors problem-task pair, with each pair of images being from one student. As can be seen in these examples, many students drew the mirrors multiple times for both the MC problem and the diagramming task.}
        \label{mirrorsdiagrams}
\end{figure}

The \mirrors\ problem-task pair asked students to consider a single beam of light that reflects off two adjacent plane-mirrors with a given angle between the mirrors. Every student drew unprompted diagrams for this problem, with many students electing to sketch the mirrors multiple times (\textit{e.g.}, Fig.~\ref{mirrorsdiagrams}(a)~left and (b)~left). The incoming and outgoing light rays cross in the MC problem, with the problem asking students to find at what angle they intersect. However, as the prompt is ambiguous as to which of these supplementary angles ($60^\circ$ or $120^\circ$) students were supposed to find, we categorized students' answers correct if they successfully identified either of these angles. While the problem statement did not give the angle at which the incoming ray hit the first mirror, no student asked for this information or even commented on it during problem solving.

Eight students answered the problem correctly, another 8 answered incorrectly, and 3 students drew unprompted diagrams without selecting an answer. For the most part, there seemed to be little correlation between the amount of detail in the unprompted diagrams and getting the correct answer, with the exception that the 4 students who selected the incorrect answer ``Not enough information'' drew very little, perhaps because they believed that adding further detail would not be productive. Interestingly, the 3 students who did not select an answer drew some of the most detailed unprompted diagrams.

We believe most students who drew the situation multiple times first started with a sketch to orient themselves to the problem, and once they realized more fully what the problem was asking, they created new, often larger, often neater sketches that they developed into ray diagrams.  Additionally, we found no correlation between students who selected the right or wrong answer with students who did or did not struggle to draw the similar situation when prompted in the diagramming task. This indicates that drawing this particular ray diagram is both challenging to do and not sufficient (by itself) to answer the \mirrors\ problem correctly. 

% What was typically drawn
% Answers
% Comparison between unprompted and prompted

\subsubsection{\efield}

\begin{figure}[t]
    \centering
        \includegraphics[width=0.406\linewidth]{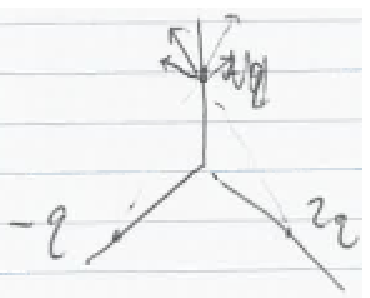}
        \includegraphics[width=0.514\linewidth]{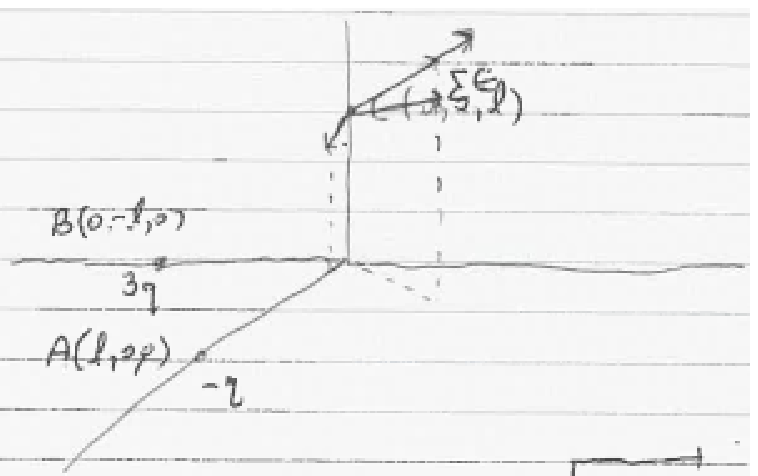}\\
        (a) \efield\ diagrams for one student.\\
        \includegraphics[width=0.396\linewidth]{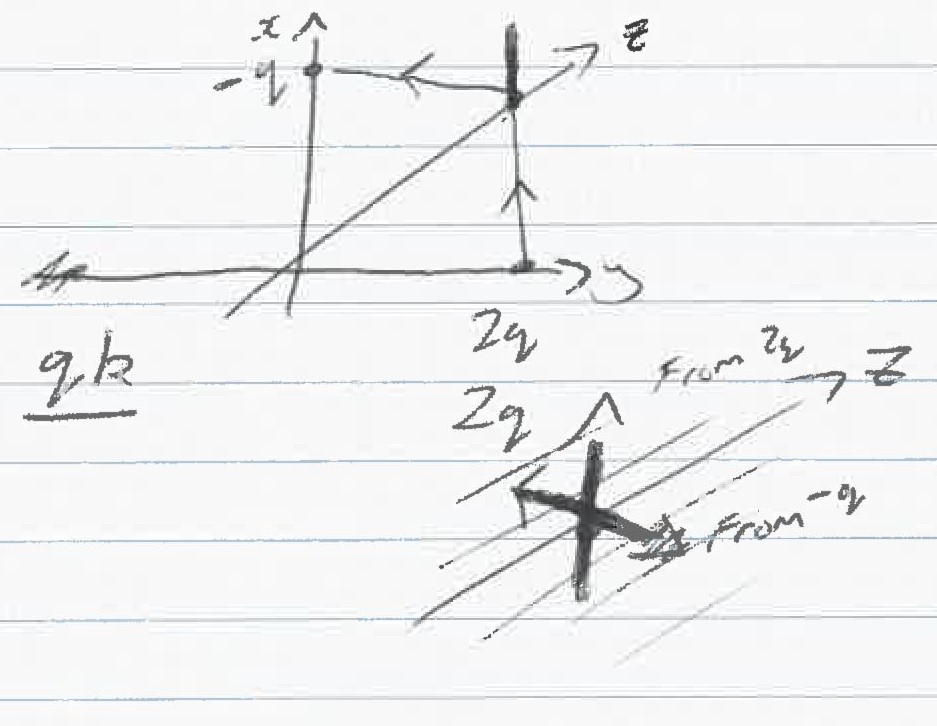}
        \includegraphics[width=0.524\linewidth]{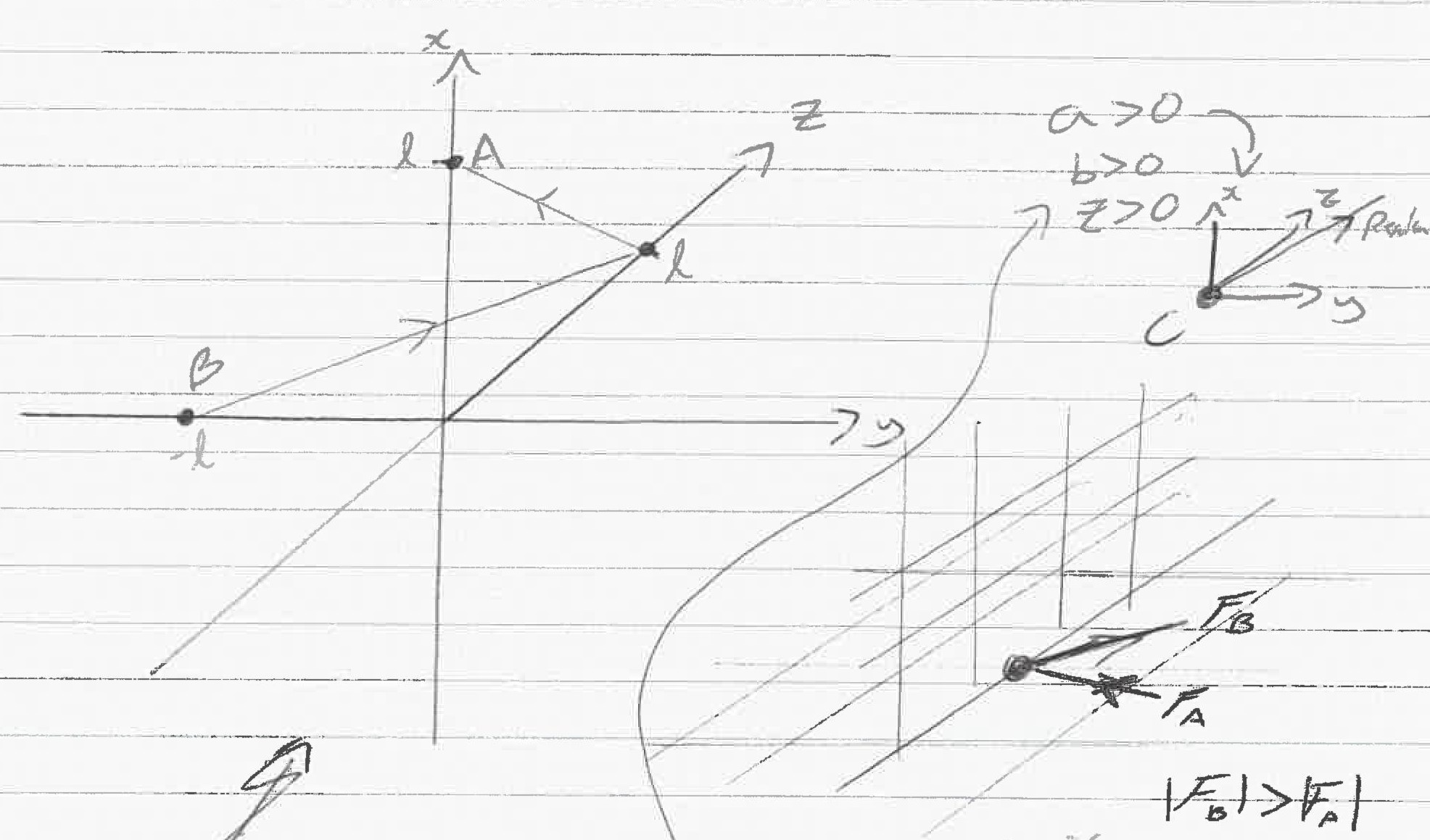}\\
        (b) \efield\ diagrams for another student.
        \caption{Examples of unprompted (left) and prompted (right) student diagrams for the maps problem-task pair, with each pair of images being from one student.}
        \label{efielddiagrams}
\end{figure}

The \efield\ scenarios asked students to consider the electric field (both magnitude and direction) at a point noncollinear with two point charges (the charges and point of interest all lie a distance $\ell$ from the origin, each on a different axis). Six students selected the correct answer for the electric field, and 5 students selected an answer with the correct magnitude and form but a sign error. None of the 5 students with a sign error labeled their axes or in anyway depicted positive or negative directions on their diagrams, whereas 4 of 7 students who did not have a sign error (the six who selected the correct answer and 1 who had only a magnitude error) labeled axes or directions. Two of the 5 students with a sign error had an accurate diagram but without direction labels, and 3 did not have a diagram or did not draw the components of the electric field. Three other students began to work on the problem but did not select an answer, and the remaining 4 students skipped the problem or were asked to skip because of time concerns.

As 5 of 6 who selected an incorrect answer only had a sign error, it is especially noteworthy that 3 of the 5 students with a sign error (who had not indicated direction in their unprompted diagrams) correctly drew the direction of the electric field and included directional labels in the prompted diagram (\textit{e.g.}, the student whose work is shown in Fig.~\ref{efielddiagrams}(a)). So while it seems that directional labels are generally necessary to solve this type of problem correctly, many students did not recognize this or chose not to include direction in their diagram without prompting. We believe that this was either an issue of activation or choice, rather than diagramming ability, since these students were largely capable of drawing accurate diagrams when prompted in the paired diagramming task. 

% What was typically drawn
% Answers
% Comparison between unprompted and prompted

\subsubsection{\deltas}
\label{deltas}

\begin{figure}[t]
    \centering
        \includegraphics[width=0.460\linewidth]{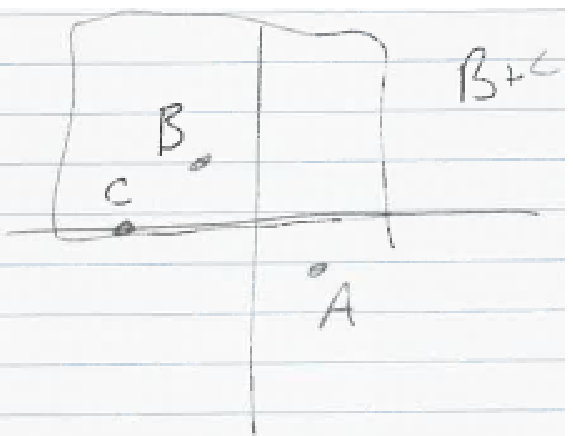}
        \includegraphics[width=0.460\linewidth]{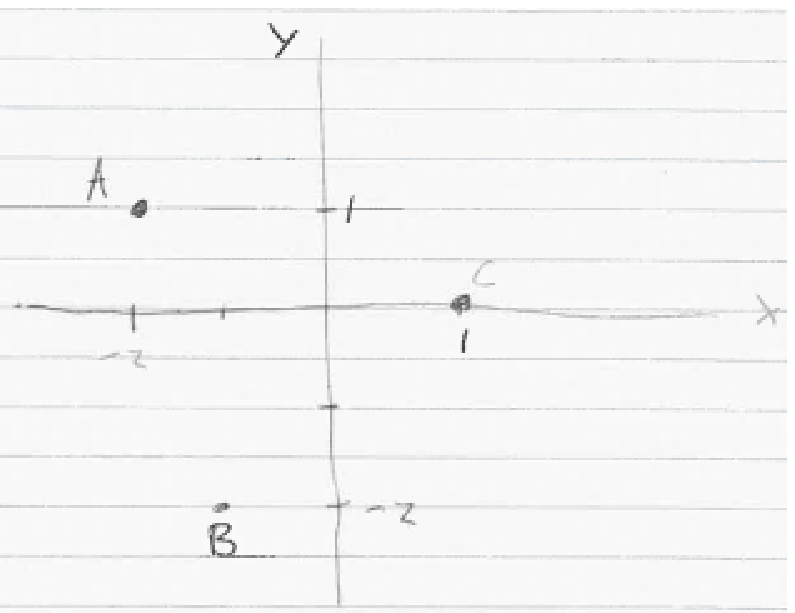}\\
        (a) \deltas\ diagrams for one student.\\
        \includegraphics[width=0.530\linewidth]{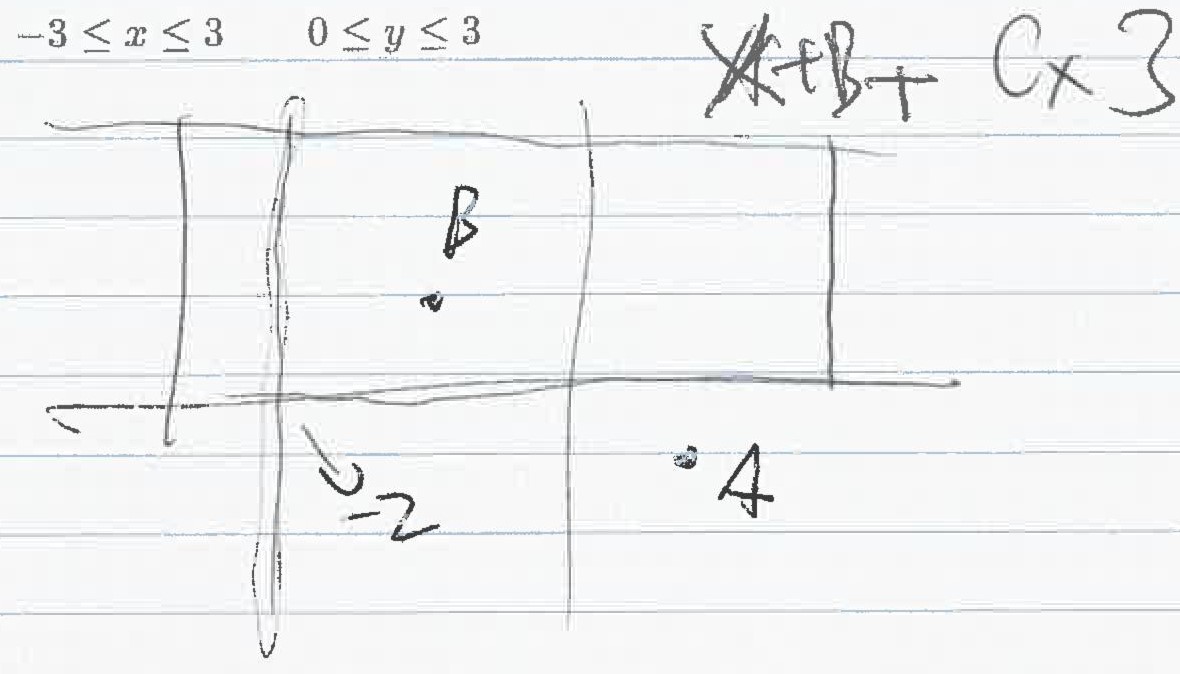}
        \includegraphics[width=0.390\linewidth]{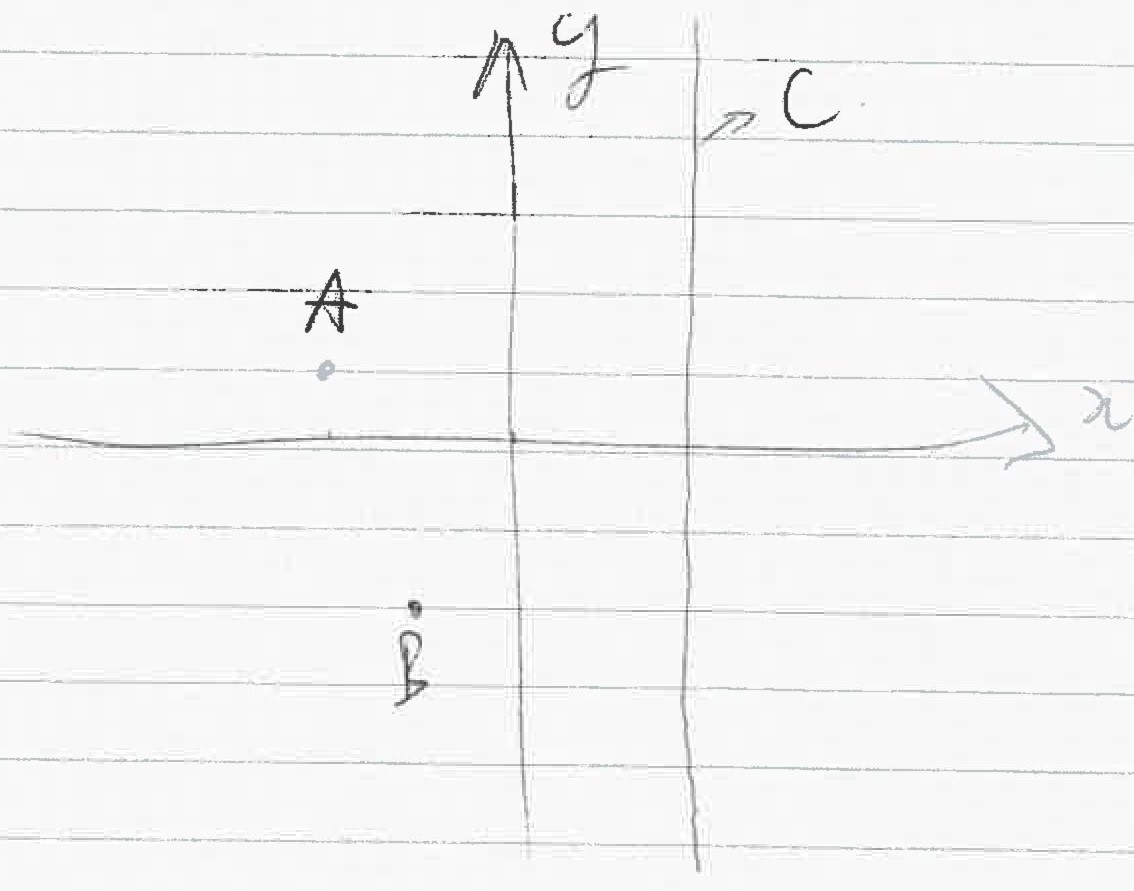}\\
        (b) \deltas\ diagrams for another student.
        \caption{Examples of unprompted (left) and prompted (right) student diagrams for the maps problem-task pair, with each pair of images being from one student.}
        \label{deltasdiagrams}
\end{figure}

The \deltas\ problem-task pair asked students to consider a 2-dimensional charge distribution ($\sigma$) containing delta functions to represent two point charges ($A$ \& $B$) and a line charge ($C$). Only 3 of 18 students correctly answered the problem (which asked how much charge is in an enclosed region of space), and only 1 of these 3 students drew an accurate unprompted diagram (in total, 8 students drew diagrams, only 2 of which were accurate). Six students stated or implied that the line charge $C$ was a point, and 4 drew it as such (\textit{e.g.}, Fig.~\ref{deltasdiagrams}). Five students, four of whom indicated $C$ was a point charge, asserted the total charge in the enclosed region was $B + C$, which is especially noteworthy as $B + C$ was not a provided option (though ``None of the above'' was an option).

One of the students who drew $C$ as a point in the MC problem \hide{16} drew $C$ correctly as a line for the paired task, indicating that they might have simply misread the expression for $\sigma$ in the problem. Eleven students did not complete the diagramming task, as it was the last item in the interview (though only 1 student was asked by the interviewer to skip this task) and it was the only task asking students to depict a scenario that would not show up in an introductory physics course. However, the high number of students who selected an incorrect answer for the problem and drew a diagram consistent with their incorrect answer suggests that many students who did not complete the diagramming task would not have been able to draw an accurate diagram given more time. The difficulties we observed with this problem-task pair are consistent with research indicating that graphical interpretations of delta functions are challenging even for upper-division physics students~\cite{wilcox_student_2015}. 

\subsection{Unpaired Multiple-Choice Problems}
\label{unpairedmcproblems}

There were 12 problems for which students may have drawn unprompted diagrams that where were not paired with a diagramming task. We now briefly discuss these problems by grouping them based on content area and patterns in the unprompted diagrams the students drew. A summary of student performance on these problems (as well as the paired MC problems) can be found in Table \ref{unpairedperformance}.

\begin{table}[t]\small
    \centering
    \caption{Student performance on all MC problems, with paired MC problems bolded. Numbers (generally $N=19$) are number of students. Columns indicate correctness of student answers to problems. Sub-columns categorize \textit{unprompted} diagram content: more than given information~(G+); only given information~(G); no diagram~(ND); or blank~(B) if no work was shown. Numbers for the paired MC problems represent a collapsing of the values in Table \ref{pairedperformance}. Students were asked to skip problems for time in the Moment of Inertia, Snell's Law, and E-potential unpaired MC problems, as well as the Decay, Deltas, and E-Field paired MC problems.}
    \label{unpairedperformance}
    \begin{tabularx}{\columnwidth}{lcccccccccccccc}\hline\hline
& \multicolumn{4}{c}{Correct} && \multicolumn{4}{c}{Incorrect} && \multicolumn{4}{c}{No Answer}\\
                        &G+&G &ND&B &&G+&G &ND&B &&G+&G &ND&B\\\hline
Two Cars                        &1 &1 &17&- &&- &- &- &- &&- &- &- &- \\
\textbf{Maps}                   &12&4 &- &- &&2 &- &- &- &&1 &- &- &- \\
Projectile                      &- &1 &14&- &&- &- &2 &- &&- &- &1 &1 \\
\textbf{Blocks}                 &9 &4 &- &- &&1 &2 &- &- &&1 &2 &- &- \\
Rolling Disk                    &2 &5 &2 &1 &&2 &2 &3 &- &&1 &- &1 &- \\
Stage Ramps                     &2 &4 &1 &2 &&1 &4 &1 &3 &&1 &- &- &- \\
Center of Mass\hspace{-2pt}     &1 &7 &- &- &&5 &3 &- &- &&- &3 &- &- \\
Collision                       &2 &6 &6 &- &&- &3 &2 &- &&- &- &- &- \\
Circular Track\hspace{-2pt}     &- &8 &8 &- &&- &- &- &- &&- &3 &- &- \\
Torque                          &12&2 &- &- &&2 &- &- &- &&2 &1 &- &- \\
M. of Inertia                   &5 &1 &6 &- &&- &- &- &- &&2 &- &3 &1 \\
\textbf{Decay}                  &2 &5 &3 &3 &&- &1 &1 &1 &&- &- &- &1 \\
\textbf{Mirrors}                &6 &2 &- &- &&4 &4 &- &- &&3 &- &- &- \\
Thin Lens                       &2 &4 &2 &- &&4 &2 &- &1 &&- &1 &- &3 \\
Snell's Law                     &11&- &1 &- &&3 &- &2 &- &&1 &- &- &- \\
\textbf{Deltas}                 &- &2 &1 &- &&- &5 &4 &2 &&- &1 &1 &2 \\
E-Potential                     &3 &2 &6 &1 &&5 &- &- &- &&- &- &- &- \\
\textbf{E-field}                 &5 &1 &- &- &&3 &2 &1 &- &&2 &- &1 &- \\ \hline\hline
    \end{tabularx}
\end{table}

\subsubsection{Two Cars and E-Potential}
\label{carsandpotential}

Two of the problems, The Two Cars and E-Potential, are isomorphic, requiring students to take the ratio of scalar values given in the prompts and then sum these ratios. Despite this mathematical similarity, all students answered the Two Cars problem correctly but only 12 of 17 students answered the E-Potential problem correctly (and 2 students were asked to skip this problem because of time constraints). Only 2 students drew diagrams for the Two Cars problem, while 10 students did so for the E-Potential problem, including the 5 who selected incorrect answer, all of whom selected ``not enough information.''

From this answer selection of ``not enough information,'' their unprompted diagrams, and some student comments, we believe these 5 students thought the positions of the charges (not just their distance from the point of interest) impacted the value of the electric potential. This dependency on position would be true for the value of the electric field at the point of interest, so it is possible students were conflating electric potential and electric field.

There are multiple possible explanations for these E-Potential diagrams: students who did not know how to solve the problem may have started by drawing the situation; students who thought the position of the charges mattered began to draw a diagram to organize this information; and/or students started to draw a diagram, but as the problem prompt did not list the positions of the charges, trying and being unable to draw the diagram may have lead students to an incorrect conclusion. 

\subsubsection{Projectiles}

Fifteen of 19 students answered the Projectiles problem correctly. Only 1 student drew an unprompted diagram---an upside-down parabola with no axes, labels, or annotations---and this student selected the correct answer. Three of the 4 students who did not select the correct answer set $y(x)=0$ rather than $y'(x)=0$ when solving for the $x$-position where the projectile begins to fall. As with the \decay\ problem discussed earlier, this issue seems to be an instance in which, while the answer could be obtained directly from a carefully drawn diagram (a graph of the given function), drawing a diagram was not necessary for students to answer the problem correctly.

\subsubsection{Rolling Disk and Moment of Inertia}

For the Rolling Disk and Moment of Inertia problems, the majority of students applied the correct physical principles, but a sizeable fraction of students struggled with proper execution. For the Rolling Disk problem, 4 of the 17 students who tried to solve the problem using conservation of energy did not take into account all three forms of energy: linear kinetic, rotational kinetic, and potential energy. For the Moment of Inertia problem, 4 of the 15 students who tried to use moment of inertia did not set up an appropriate integral. In addition, for both problems, a number of students mis-remembered formulas (\textit{e.g.}, $\omega = vr$) and/or made algebraic mistakes while solving the problems.

For the Rolling Disk problem, 7 of 12 students who drew a diagram selected the correct answer, while 5 of 9 who did not draw a diagram selected the correct answer. For the Moment of Inertia problem, 5 of 7 students who drew a diagram selected the correct answer, while 6 of 10 who did not draw a diagram selected the correct answer. For these problems, it seems like students who drew diagrams were slightly more likely to answer the question correctly, but this does not mean that the diagram helped the students answer correctly. As may be the case with the \deltas\ problem (Sec.~\ref{deltas}), it is possible that students who had a better grasp of these situations were simply more able to draw them, and it did not seem as though drawing diagrams for these problems helped students catch mistakes that they had made, either in setting up their equations or in their algebra.

\subsubsection{Center of Mass and Torque}

The Center of Mass and Torque problems both required students to take into account multiple ``real world'' objects (not just ``masses'' or ``charges''). Fifteen students used the center of mass equation, but only 7 of those got the correct answer. Six students did not take into account the mass of the tray on which the items sat for the center of mass problem, either because they forgot it or because they intentionally---and erroneously---excluded it because it's center of mass was at the origin. Two other students solved for just the x position of the COM and ignored both the tray and the plate, possibly because they lie on the y axis. One student got the correct answer by, we gather from verbal utterances, sensemaking about the COM being closer to heavier objects.

Eleven students correctly applied the principals of torque to answer the Torque question correctly, while one student correctly used proportional reasoning and superposition. Two students made mistakes while trying to use torque, and 4 students only used or calculated forces while trying to solve this problem. 

Every student drew an unprompted diagram for both of these problems. As with the mirrors problem, the diagrams drawn by students who did not select an answer contained the largest number of details. Along with other analyses (discussed in Sec. \ref{quant}), this suggests that in these instances, students tried to use diagrams to help reach an answer they were unable to obtain through other means (\textit{e.g.}, through algebra). Additionally, many students who did not select the correct answer were able to generate accurate unprompted diagrams, which again parallels the Mirrors problem. 

\subsubsection{Stage Ramps}

Although friction was not discussed in the Stage Ramps problem statement, students needed to consider friction in order to answer the Stage Ramps question correctly (which 9 students did). Twelve students drew diagrams for the Stage Ramps problem, including 6 of the 9 who considered friction. Only one student included a sketch that depicted a friction force. As we saw in the \mirrors, Center of Mass, and Torque problems, the most detailed diagrams were drawn by students who did not select an answer.

\subsubsection{Collision}

All students used conservation of momentum in the Collision problem, but 5 students had a sign error. Two of these students set up the equations with an incorrect sign, and 3 students had the correct equations but made a sign error while solving for the answer. The other 14 students all got the correct answer.

Eleven students drew unprompted diagrams for this problem, including 8 who got the correct answer. Three of 5 students with a sign error drew an (accurate) unprompted diagram. One student who drew a diagram, and two students who did not, set up their conservation of momentum equation with a sign error: the other two students who selected the wrong answer made an algebraic mistake. Overall, 10 of 11 students who drew a diagram set up their equations correctly, so it is possible (as with the \efield\ problem) that drawing a diagram that depicts direction can help students avoid sign errors. 

\subsubsection{Circular Track}

Ten students drew diagrams for the Circular Track problem (one of whom showed no other work). Eighteen students used rotational kinematics to solve the problem, with 16 students getting the correct answer and three students, including a student who only showed a sketch, not selecting an answer. Seven of the 16 students who selected the correct answer drew a diagram. As with the \mirrors, Center of Mass, Torque, and Stage Ramps problems, the most detailed diagrams were drawn by students who did not select an answer.

\subsubsection{Thin Lens and Snell's Law}

To find the magnification of an image from a thin lens, 13 students drew diagrams: 6 got the right answer (2 of whom got the answer from just the diagram), 6 got the wrong answer, and 1 did not select an answer. Two students got the correct answer without drawing any diagrams. The diagrams drawn for this problem are similar to those drawn for the \decay\ problem in that many of the diagrams lacked the features that would have been necessary to aid in solving the problem from the diagram. 

One student was asked to skip the Snell's Law problem and four students only provided a sketch (one of whom selected the correct answer). All 14 other students used Snell's Law, with 11 getting the answer correct and 3 making minor mistakes leading to an incorrect answer. Of students who drew a diagram, 6 of 11 who got the correct answer never drew the refracted rays of light, as was the case with 1 of the 3 students with a diagram who selected an incorrect answer.

It seems that many of these diagrams were drawn before the students switched to a completely algebraic mode of problem solving that did not depend on the diagram, similar to the \decay\ problem.

\section{Quantitative Results and Interpretations}
\label{quant}

As discussed in Sec. \ref{sec:Methods}, we qualitatively coded every detail that appeared on student diagrams. An illustrative example of this coding is shown in Fig.~\ref{coded_diagram}. We would also like to remind the reader that all of the students who participated in our study were physics majors who had completed at least a full year of physics, and thus it is likely that some of our findings may not generalize to physics majors in introductory courses or non-physics majors in physics courses.

\begin{figure}[t]
    \centering
        \includegraphics[width=\linewidth]{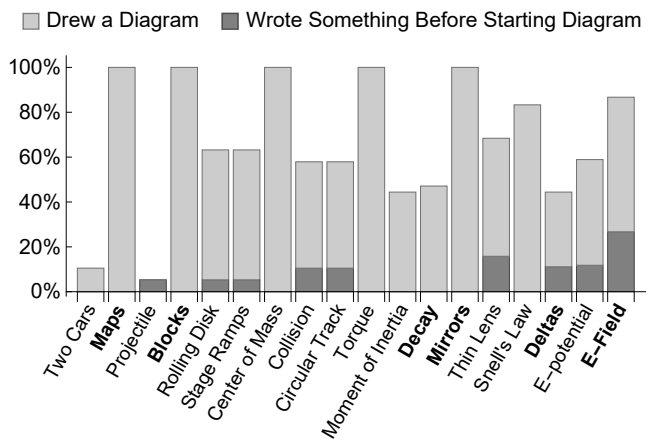}
    \caption{The percentage of students who drew a diagram for each MC problem. The darker shading indicates instances in which a diagram was drawn but not the first thing the student put onto the paper.}
    \label{diagramsfirst}
\end{figure}

In total, across the 19 students and 18 problems, students worked through 331 problems (with 11 instances where we asked a student to skip a problem due to time constraints). Students drew diagrams for 66.2\% of these problems ($N=219$), and Fig.~\ref{diagramsfirst} shows what fraction of students drew a diagram for each problem prompt. We identified a total of 1771 diagram details, including 1051 markings (\textit{e.g.}, arrows, objects, etc.), 644 (non-axis) labels (141, or 21.9\%, of which had units), and 76 axis details (drawing or labeling axes). Since student use of axes was not a primary focus of this study, the coding of axes details was not as granular as the coding of other diagram details: for example, drawing both an $x$- and $y$-axis would be coded as a single axes detail, which is part of why the number of axis details appear to represent such a small portion of the overall details.

Also shown in Fig.~\ref{diagramsfirst} is that there were very few instances---just 18 of 219---in which students drew a diagram but wrote something else down first. In these instances, students typically were doing some other form of orienting (writing down general formulas, rewriting key information from the prompt, or identifying relevant physical laws, etc.), and the diagram quickly followed. There were only two instances in which it appears a student tried to solve a problem without a diagram, then after being unsuccessful drew a diagram as part of a second attempt. With 91.8\% of diagrams being the first thing students drew, and with diagrams coming very early in the problem-solving process for the other 18 instances, we conclude that most, if not all, unprompted diagrams drawn in this study helped students orient themselves to the problem they were trying to solve---however, as we discuss below, we believe many of these diagrams served additional functions as well.

The finding that 91.8\% of unprompted diagrams were started before anything else was written down came from recording the timing of each of the 1771 details that we had identified and qualitatively coded. This time-stamping was done to better understand when and how students used diagrams for orienting and for other purposes. In addition, every problem was coded along multiple axes: Student, Problem, Answer (Correct, Incorrect, Blank), Problem-solving Duration, Diagram (Diagram, No Diagram), etc. If a student drew an unprompted diagram while solving a problem, we also coded: the number of details in the diagram, and whether something was written before the student started the diagram (Fig. \ref{diagramsfirst}). Every diagram detail was coded as given or not given in the problem statement. We also added a code for the cohort the student was in: lower-division (undergraduate), junior, senior, or graduate student. We did not code or timestamp algebra or other actions taken by students (\textit{e.g.}, using a calculator, speaking, etc.), other than the first non-diagram action taken after the diagram was started. 

\begin{figure}[t]
    \centering
        \includegraphics[width=\linewidth]{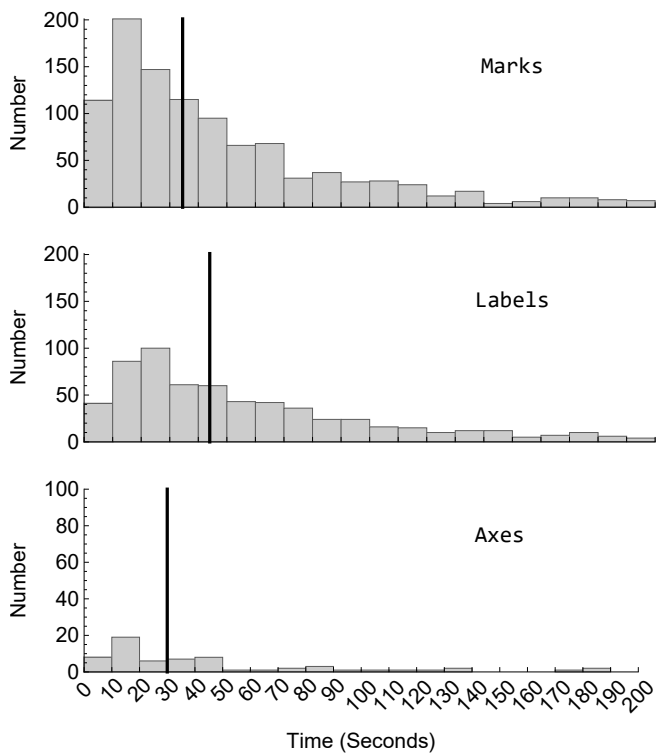}
    \caption{Histogram of student diagram details (marks, labels, and axes) by 10 second intervals for 219 unprompted diagrams. For example, between 10 and 20 seconds after starting a problem, students put a total of 201 markings (\textit{e.g.}, arrows, objects, etc.), 86 non-axis labels, and 19 axis details (axes or axis labels) on diagrams. These graphs includes 97\% (n=1705) of diagram details, with the remaining 3\% (n=46) occurring between 200 and 598 seconds. Vertical bars indicate median times. Note: The vertical axis for the Axes plot is different because of the low number of axes details.}
    \label{time(s)}
\end{figure}

\begin{figure}[t]
    \centering
        \includegraphics[width=\linewidth]{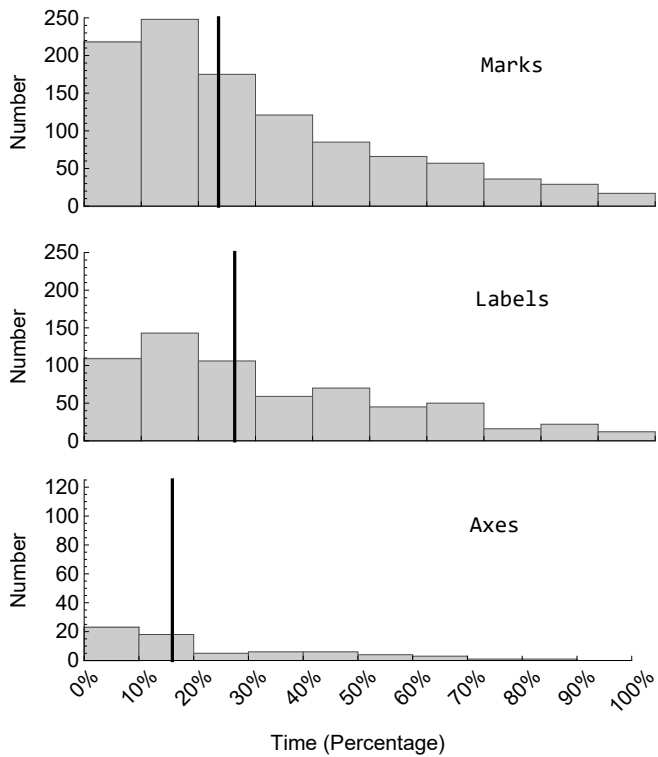}
    \caption{Histogram of timing of student diagram details (marks, labels, and axes) as a fraction of total time spent on each problem. For example, a total of 218 markings, 109 non-axis labels, and 23 axis details were drawn by students in the first 10\% of time spent problem solving. Vertical bars indicate median times. Note: The vertical axis for the Axes plot is different because of the low number of axes details.}
    \label{time(p)}
\end{figure}

The overall results of time-stamping the details in unprompted diagrams is displayed in Figs.~\ref{time(s)} and~\ref{time(p)}, which show how long after starting the problem the students made various types of markings (both in absolute time---seconds---and as a fraction of the total time the student spent on the problem). Both in absolute and relative terms, the vast majority of diagramming occurs near the start of problem-solving process. As a whole, median time of the last diagram detail was 55 seconds or 45\% of the way through solving the problem.

With 19 students completing a total of 331 problems, drawing 219 unprompted diagrams and a total of 1771 markings, we are able to look at patterns in when and how the students generated and used unprompted diagrams. When comparing two categorical variables---where our categorical variables include Student, Problem, Cohort, Answer, or if a Diagram was drawn---we used a Fisher Exact (FE) test to determine statistical significance, where p-values were simulated via Monte Carlo. When comparing a categorical value with a numerical value (Duration, Number of Details), we used a Kruskal-Wallis (KW) test. These statistical tests were run in the programming language R.

When considering a large number of statistical tests, as we did in this study, it is important to account for the possibility of false positives. Since a p-value of 0.05 signifies a 5\% chance of the observed result being being seen if the null hypothesis is true, running a large number of tests increases the likelihood that some of the results will be false positives. To avoid this pitfall, we use the conservative Bonferroni-adjusted p-value, which is calculated by dividing the 0.05 threshold by the number of tests we run on a particular subset of the data (\textit{e.g.}, all tests looking at correlations with students selecting the correct answer). We ran no more than 10 statistical tests on any subset of our data, so we can use a (relatively conservative) p-value threshold of 0.005 for exploratory analyses. For any test that returns a p-value between 0.05 and 0.005, we only argue the result is significant if it is confirming an expected result.  In the sections below, we examine first trends that were examined to confirm an existing hypothesis and then exploratory analysis to identify any unanticipated trends in the data.  

\subsection{Confirmatory Analyses}

In this section, we discuss findings that, from our interview development and previous work~\cite{vignal_comparing_2020}, we expected to be statistically significant. Most, but not all, of these findings did prove to be significant.

We suspected that graduate students would answer the most questions correct for a number of reasons: they have had the most formal physics education, they elected to pursue more advanced physics degrees, and many graduate students teach in introductory physics courses. Indeed, we found that graduate students were the most likely group to select a correct answer (78.7\% of the time), compared to seniors (55.1\%), juniors (55.8\%), and lower-division students (61.2\%). That graduate students answered more questions correctly than did any other cohort is statistically significant (FE~$p=0.017$), but the differences between the other groups were not.

In choosing the order of questions when designing the survey, we intentionally places the E-Potential problem before the E-Field problem. While there is enough information in the E-Potential problem to solve the problem, there is not enough information to draw an accurate diagram (since the coordinates of the charges are not know). As the E-Fields problem requires students to consider relative positions and not just distances, we did not want to cue students to think about coordinates in the E-Potential problem by having it come after the E-Fields problem. Such cuing, we argue, could bias students towards both generating diagrams that they otherwise might not have drawn and then concluding that there was not enough information to solve the problem. From this conversation, we wondered if students who generated unprompted diagrams for the E-Potential problem would be less likely to select a correct answer than those who did not. This ended up being the case (FE~p=0.044), where 100\% of students (7 of 7) without a diagram selected the correct answer, compared to 50\% (5 of 10) of students who drew a diagram. As discussed briefly in Sec. \ref{carsandpotential}, it could be that students who did not know how to solve the problem were more likely to generate a diagram for it, or that the process of generating a diagram hindered students ability to solve the problem.

In our initial study of the 6 MC problems that were paired with diagramming task, we observed that graduate students were the most likely to draw unprompted diagrams and that lower-division students were the least likely to draw unprompted diagrams. By looking at the 12 MC problems that were not paired with a diagramming task, we are able to test these findings. Across all 18 MC problems, graduate students drew diagrams 74.2\% of the time, compared to 65.2\% for seniors, 68.6\% for juniors, and 53.7\% for lower-division students. The difference between lower-division students and other students is statistically significant (FE~p=0.043), but the difference between graduate students and other students is not (FE~p=0.065). For this final result, we want to note that this p-value does not disprove a correlation, it merely means that we cannot be confident (with our data set) that the null hypothesis is false. It is possible that a larger data set would have the statistical power to identify this trends as significant.

Finally, in our initial study of the problem-task pairs, we had also found that lower-division students, while they drew the fewest unprompted diagrams, tended to include the highest number of details in those diagrams. Testing this observation using the 12 unpaired MC problems, we find that the difference in number of details between the groups is not statistically significant (KW~p=0.353).

\subsection{Exploratory Analyses}

As can be seen in Fig.~\ref{diagramsfirst}, between 5.3\% and 100\% of students drew diagrams on any given problem, and the correlation between the specific problem being solved and the fraction of students who drew unprompted diagrams was statistically significant (FE~p$<$0.001). This result is somewhat surprising, given that most of these interview questions were designed to evoke unprompted diagrams in students. 

Overall, the presence of a diagram was not a significant predictor of whether a student got the answer correct (FE~p=0.365) when looking across all 18 MC problems (though, as mentioned above, it was significant in the case of the E-Potential problem). This result makes sense with our previous findings that, for some problems, diagrams do not seem to be necessary in order for students to select the correct answer, and for other problems, a large number of students who drew diagrams were still unable to select the correct answer.

In addition to certain problems evoking student diagrams more or less frequently, different students were more or less likely to draw unprompted diagrams. The percentage of problems for which individual students drew a diagram range from 33.3\% to 94.7\%, a correlation that was statistically significant (FE~p$<$0.001). Considering this result in light of our findings that, as cohorts, only lower-division students were statistically distinct, we believe this shows that the variation within each cohort of students is greater than the variation between the cohorts.

For 122 of the problems in which a student drew an unprompted diagram (55.7\% of the time), students added details to their diagram that were not given in the problem statement. These details were significantly correlated with which problem a student was solving (FE~p$<$0.001), but were independent of which student (or cohort) was solving the problem and whether students selected the correct answer. From this we conclude that the extent to which a student used their diagram to organize information depends on the problem but does not predict success. 

In 102 instances (46.6\% of times when a diagram was drawn), a student revisited the diagrams to add to it (or created additional diagrams) after doing something else (generally performing algebra). This was associated with a higher likely-hood of not getting the answer right---selecting an incorrect answer or no answer---with 72\% of non-revisited diagrams and 52\% of revisited diagrams being drawn on diagrams with the correct answer selected (FE~p=0.003). Revisiting a problem is highly correlated with which problem the student was trying to solve (FE~p$<$0.001), but was not correlated with the student or group of students. From this, we do not conclude that revisiting a diagram is detrimental to students, rather that it seems most likely that students would revisit a diagram while attempting to solve a challenging problem.

In our initial design of these interview problems, we wanted to present students with problems that were possible to solve entirely from a diagram, and we believe 6 of the problems could be reasonably solved in a purely geometric way. This happened rarely. There were only 9 times students obtained an answer directly from a diagram, 5 times for the Deltas problem (1 correct answer, 2 incorrect answers, and 2 students who did not select an answer because the answer they obtained was not listed as an answer option.) and 3 times for the Thin Lens problem (2 correct answers, 1 incorrect answer). As we posited previously, this finding suggests that there may be room to improve or increase instruction around geometric solutions in physics~\cite{vignal_comparing_2020}.

When looking at how long it took students to solve problems, we found the largest difference was that that lower-division students spent on average ~9\% more time solving problems than did students in the other cohorts, though this finding was not statistically significant (KW~p=0.065).

\begin{figure}[t]
    \centering
        \includegraphics[width=\linewidth]{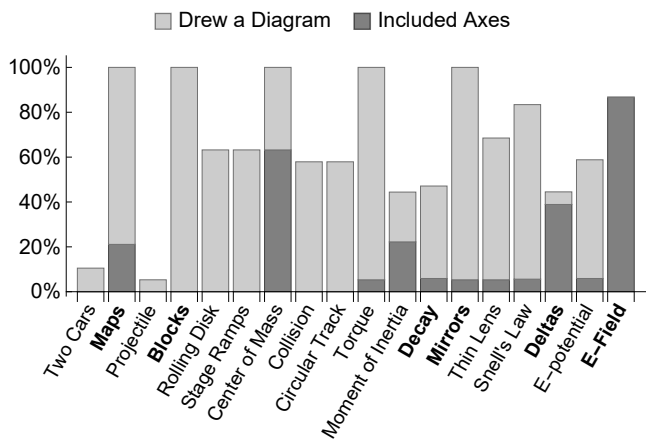}
    \caption{The percent of diagrams drawn (for each problem prompt) that included axes. Axes were most common for the problems that directly (Center of Mass, E-Potential) or indirectly (Deltas) provided coordinate information in the problem prompt.}
    \label{diagramswithaxes}
\end{figure}

Finally, while looking at student use of axes was not a primary focus of this work, we did find that students drew axes on only 21.0\% of unprompted diagrams (46 of 219), with no individual student drawing axes for more than 6 problems. Fig.~\ref{diagramswithaxes} shows the percentage of student-generated diagrams for each problem that included axes. As the problems for which students drew axes the most were problems that gave students information (directly or indirectly) in the form of coordinates, these findings are not surprising. These findings do, however, allow us to nuance our previous claim that students are much about twice as likely to add axes to prompted diagrams than to unprompted diagrams: this discrepancy---which is not necessarily bad and may reflect reasonable decisions on the part of students---may be even larger problems with non-coordinate information.  

% %--------------------------------- DISCUSSION ----------------------------
\section{Synthesis and Conclusion}
\label{sec:Discussion}
In this study, we presented 19 physics majors with 18 multiple-choice physics problems and 6 diagramming tasks, where each task was of a similar physical situation as 1 of the multiple-choice problems. The 18 multiple-choice problems gave us the opportunity to student diagrams that students generate without prompting during the problem-solving process. The paired diagramming tasks then let us compare prompted and unprompted diagrams in 6 different physical contexts.

Previous work looking specifically at the problem-task pairs has been published~\cite{vignal_comparing_2020}. We found that prompted diagrams were generally larger, objects in the diagrams were more to scale, and these diagrams were more likely to include axes and units, suggesting that these features are more valued in diagrams created for the purpose of communicating information than they are in diagrams generated to orient to a physical scenario and/or diagrams generated as a tool to manipulate throughout the problem-solving process.

This previous work also found instances in which students selected the correct answer without drawing a diagram or selected an incorrect answer after drawing a diagram, a finding confirmed in this study when looking at a separate subset of the multiple-choice problems. Thus, we reiterate our previous suggestion that instructors consider when and how to teach and assess diagrams in problem solving, as the problem context seems to matter (and in some cases, requiring students to draw diagrams may negatively impact their ability to select a correct answer~\cite{heckler_consequences_2010}). This is not to say that diagramming should not be taught in these contexts, merely that there should be alignment between the goals of the course and assessment when it comes to diagramming. 

In this study, we found that the vast majority, if not all, of unprompted diagrams generated by students were used (at least in part) to help students orient themselves to the problem. An additional study already underway, in which we present students and faculty with a modified subset of these interview questions and then explicitly ask them about the role of diagrams in problem solving, will seek to confirm and elaborate on this finding. In addition, using eye-tracking software during this study could help us more fully understand when and how participants refer back to their unprompted diagrams~\cite{susac_role_2019}.

Furthermore, having coded 1771 diagram details for 219 multiple-choice problems, we are confident in asserting that problem-solving diagrams are messy. In addition to considering how to value messy diagrams in assessment, we believe it is important for instructors to model messy diagrams for students. The differences between these messy diagrams and carefully constructed figures in textbooks and lectures are stark: not only are these diagrams generated by different people with different levels of experience, they are generated for different purposes. Yet if students are never exposed to examples of messy work (including messy diagrams) done by experts, and if students' messy work is not valued in assessment, then students may come to believe that the messy work they are doing is not good physics.

Our study is limited in that all students interviewed were physics majors at the University of Colorado Boulder, which has a strong undergraduate physics program and highly-selective graduate physics program. Furthermore, the student demographics of this department are overwhelmingly white and male. Our findings may not generalize to other populations, and so further study of student diagramming, especially with non-physics majors or physics majors in introductory courses, may be necessary.

We believe there are many affordances to our study design: by having students solve problems without specific prompts (diagramming or otherwise), and then by giving students explicit prompts, this study design can be used to explore other areas of student problem-solving and how tools used in problem-solving manifest across different epistemic frames. One change our study design may have benefited from would have been setting aside time at the end of the interview to explicitly discuss diagramming with students, and this modification is the primary focus of ongoing work. Together with this current study, this work can help inform further work into student-generated diagrams, as well as research and instructional approaches regarding student interactions with expert-generated diagrams.

% %--------------------------------- CONCLUSIONS ----------------------------

% \section{Conclusions}
% \label{sec:Conclusions}

% %--------------------------------- ACKNOWLEDGEMENTS ----------------------------
\section*{Acknowledgements}
We would like to acknowledge members of the PER@C group, including our pilot interviewees, as well as the students who participated in the interviews. This work was supported by the University of Colorado Boulder Department of Physics.

% %--------------------------------- BIBLIOGRAPHY ----------------------------

% \
% \bibliographystyle{apsrev4-1}
\bibliography{bib.bib}

%apsrev4-2.bst 2019-01-14 (MD) hand-edited version of apsrev4-1.bst
%Control: key (0)
%Control: author (8) initials jnrlst
%Control: editor formatted (1) identically to author
%Control: production of article title (0) allowed
%Control: page (0) single
%Control: year (1) truncated
%Control: production of eprint (0) enabled
\begin{thebibliography}{30}%
\makeatletter
\providecommand \@ifxundefined [1]{%
 \@ifx{#1\undefined}
}%
\providecommand \@ifnum [1]{%
 \ifnum #1\expandafter \@firstoftwo
 \else \expandafter \@secondoftwo
 \fi
}%
\providecommand \@ifx [1]{%
 \ifx #1\expandafter \@firstoftwo
 \else \expandafter \@secondoftwo
 \fi
}%
\providecommand \natexlab [1]{#1}%
\providecommand \enquote  [1]{``#1''}%
\providecommand \bibnamefont  [1]{#1}%
\providecommand \bibfnamefont [1]{#1}%
\providecommand \citenamefont [1]{#1}%
\providecommand \href@noop [0]{\@secondoftwo}%
\providecommand \href [0]{\begingroup \@sanitize@url \@href}%
\providecommand \@href[1]{\@@startlink{#1}\@@href}%
\providecommand \@@href[1]{\endgroup#1\@@endlink}%
\providecommand \@sanitize@url [0]{\catcode `\\12\catcode `\$12\catcode
  `\&12\catcode `\#12\catcode `\^12\catcode `\_12\catcode `\%12\relax}%
\providecommand \@@startlink[1]{}%
\providecommand \@@endlink[0]{}%
\providecommand \url  [0]{\begingroup\@sanitize@url \@url }%
\providecommand \@url [1]{\endgroup\@href {#1}{\urlprefix }}%
\providecommand \urlprefix  [0]{URL }%
\providecommand \Eprint [0]{\href }%
\providecommand \doibase [0]{https://doi.org/}%
\providecommand \selectlanguage [0]{\@gobble}%
\providecommand \bibinfo  [0]{\@secondoftwo}%
\providecommand \bibfield  [0]{\@secondoftwo}%
\providecommand \translation [1]{[#1]}%
\providecommand \BibitemOpen [0]{}%
\providecommand \bibitemStop [0]{}%
\providecommand \bibitemNoStop [0]{.\EOS\space}%
\providecommand \EOS [0]{\spacefactor3000\relax}%
\providecommand \BibitemShut  [1]{\csname bibitem#1\endcsname}%
\let\auto@bib@innerbib\@empty
%</preamble>
\bibitem [{\citenamefont {Scaife}\ and\ \citenamefont
  {Rogers}(1996)}]{scaife_external_1996}%
  \BibitemOpen
  \bibfield  {author} {\bibinfo {author} {\bibfnamefont {M.}~\bibnamefont
  {Scaife}}\ and\ \bibinfo {author} {\bibfnamefont {Y.}~\bibnamefont
  {Rogers}},\ }\bibfield  {title} {\bibinfo {title} {External cognition: how do
  graphical representations work?},\ }\href
  {https://doi.org/10.1006/ijhc.1996.0048} {\bibfield  {journal} {\bibinfo
  {journal} {International Journal of Human-Computer Studies}\ }\textbf
  {\bibinfo {volume} {45}},\ \bibinfo {pages} {185} (\bibinfo {year}
  {1996})}\BibitemShut {NoStop}%
\bibitem [{\citenamefont {Foster}(2002)}]{foster_implications_2002}%
  \BibitemOpen
  \bibfield  {author} {\bibinfo {author} {\bibfnamefont {T.~M.}\ \bibnamefont
  {Foster}},\ }\bibfield  {title} {\bibinfo {title} {Implications of
  {Distributed} {Cognition} for {PER}},\ }in\ \href
  {https://www.compadre.org/portal/items/detail.cfm?ID=4372} {\emph {\bibinfo
  {booktitle} {2002 {Physics} {Education} {Research} {Conference}
  {Proceedings}}}}\ (\bibinfo {year} {2002})\BibitemShut {NoStop}%
\bibitem [{\citenamefont {Rosengrant}(2006)}]{rosengrant_case_2006}%
  \BibitemOpen
  \bibfield  {author} {\bibinfo {author} {\bibfnamefont {D.}~\bibnamefont
  {Rosengrant}},\ }\bibfield  {title} {\bibinfo {title} {Case {Study}:
  {Students}’ {Use} of {Multiple} {Representations} in {Problem} {Solving}},\
  }in\ \href {https://doi.org/10.1063/1.2177020} {\emph {\bibinfo {booktitle}
  {{AIP} {Conference} {Proceedings}}}},\ Vol.\ \bibinfo {volume} {818}\
  (\bibinfo  {publisher} {AIP},\ \bibinfo {address} {Salt Lake City, Utah
  (USA)},\ \bibinfo {year} {2006})\ pp.\ \bibinfo {pages} {49--52}\BibitemShut
  {NoStop}%
\bibitem [{\citenamefont {Kohl}\ \emph {et~al.}(2007)\citenamefont {Kohl},
  \citenamefont {Finkelstein}, \citenamefont {Hsu}, \citenamefont {Henderson},\
  and\ \citenamefont {McCullough}}]{kohl_expert_2007}%
  \BibitemOpen
  \bibfield  {author} {\bibinfo {author} {\bibfnamefont {P.~B.}\ \bibnamefont
  {Kohl}}, \bibinfo {author} {\bibfnamefont {N.~D.}\ \bibnamefont
  {Finkelstein}}, \bibinfo {author} {\bibfnamefont {L.}~\bibnamefont {Hsu}},
  \bibinfo {author} {\bibfnamefont {C.}~\bibnamefont {Henderson}},\ and\
  \bibinfo {author} {\bibfnamefont {L.}~\bibnamefont {McCullough}},\ }\bibfield
   {title} {\bibinfo {title} {Expert and {Novice} {Use} of {Multiple}
  {Representations} {During} {Physics} {Problem} {Solving}},\ }in\ \href
  {https://doi.org/10.1063/1.2820914} {\emph {\bibinfo {booktitle} {{AIP}
  {Conference} {Proceedings}}}}\ (\bibinfo  {publisher} {AIP},\ \bibinfo
  {address} {Greensboro (NC)},\ \bibinfo {year} {2007})\ pp.\ \bibinfo {pages}
  {132--135}\BibitemShut {NoStop}%
\bibitem [{\citenamefont {Heckler}(2010)}]{heckler_consequences_2010}%
  \BibitemOpen
  \bibfield  {author} {\bibinfo {author} {\bibfnamefont {A.~F.}\ \bibnamefont
  {Heckler}},\ }\bibfield  {title} {\bibinfo {title} {Some {Consequences} of
  {Prompting} {Novice} {Physics} {Students} to {Construct} {Force}
  {Diagrams}},\ }\href {https://doi.org/10.1080/09500690903199556} {\bibfield
  {journal} {\bibinfo  {journal} {International Journal of Science Education}\
  }\textbf {\bibinfo {volume} {32}},\ \bibinfo {pages} {1829} (\bibinfo {year}
  {2010})}\BibitemShut {NoStop}%
\bibitem [{\citenamefont {Kirsh}(2010)}]{kirsh_thinking_2010}%
  \BibitemOpen
  \bibfield  {author} {\bibinfo {author} {\bibfnamefont {D.}~\bibnamefont
  {Kirsh}},\ }\bibfield  {title} {\bibinfo {title} {Thinking with external
  representations},\ }\href {https://doi.org/10.1007/s00146-010-0272-8}
  {\bibfield  {journal} {\bibinfo  {journal} {AI \& SOCIETY}\ }\textbf
  {\bibinfo {volume} {25}},\ \bibinfo {pages} {441} (\bibinfo {year}
  {2010})}\BibitemShut {NoStop}%
\bibitem [{\citenamefont {Tairab}\ and\ \citenamefont
  {Al-Naqbi}(2004)}]{tairab_how_2004}%
  \BibitemOpen
  \bibfield  {author} {\bibinfo {author} {\bibfnamefont {H.~H.}\ \bibnamefont
  {Tairab}}\ and\ \bibinfo {author} {\bibfnamefont {A.~K.~K.}\ \bibnamefont
  {Al-Naqbi}},\ }\bibfield  {title} {\bibinfo {title} {How do secondary school
  science students interpret and construct scientific graphs?},\ }\href
  {https://doi.org/10.1080/00219266.2004.9655920} {\bibfield  {journal}
  {\bibinfo  {journal} {Journal of Biological Education}\ }\textbf {\bibinfo
  {volume} {38}},\ \bibinfo {pages} {127} (\bibinfo {year} {2004})}\BibitemShut
  {NoStop}%
\bibitem [{\citenamefont {Meltzer}(2005)}]{meltzer_student_2005}%
  \BibitemOpen
  \bibfield  {author} {\bibinfo {author} {\bibfnamefont {D.~E.}\ \bibnamefont
  {Meltzer}},\ }\bibfield  {title} {\bibinfo {title} {Student {Learning} {In}
  {Upper}-{Level} {Thermal} {Physics}: {Comparisons} {And} {Contrasts} {With}
  {Students} {In} {Introductory} {Courses}},\ }in\ \href
  {https://doi.org/10.1063/1.2084694} {\emph {\bibinfo {booktitle} {{AIP}
  {Conference} {Proceedings}}}},\ Vol.\ \bibinfo {volume} {790}\ (\bibinfo
  {publisher} {AIP},\ \bibinfo {address} {Sacramento, California (USA)},\
  \bibinfo {year} {2005})\ pp.\ \bibinfo {pages} {31--34}\BibitemShut {NoStop}%
\bibitem [{\citenamefont {Nguyen}\ \emph {et~al.}(2009)\citenamefont {Nguyen},
  \citenamefont {Rebello}, \citenamefont {Sabella}, \citenamefont {Henderson},\
  and\ \citenamefont {Singh}}]{nguyen_students_2009}%
  \BibitemOpen
  \bibfield  {author} {\bibinfo {author} {\bibfnamefont {D.-H.}\ \bibnamefont
  {Nguyen}}, \bibinfo {author} {\bibfnamefont {N.~S.}\ \bibnamefont {Rebello}},
  \bibinfo {author} {\bibfnamefont {M.}~\bibnamefont {Sabella}}, \bibinfo
  {author} {\bibfnamefont {C.}~\bibnamefont {Henderson}},\ and\ \bibinfo
  {author} {\bibfnamefont {C.}~\bibnamefont {Singh}},\ }\bibfield  {title}
  {\bibinfo {title} {Students’ {Difficulties} in {Transfer} of {Problem}
  {Solving} {Across} {Representations}},\ }in\ \href
  {https://doi.org/10.1063/1.3266720} {\emph {\bibinfo {booktitle} {2009
  {Physics} {Education} {Research} {Conference} {Proceedings}}}}\ (\bibinfo
  {year} {2009})\ pp.\ \bibinfo {pages} {221--224}\BibitemShut {NoStop}%
\bibitem [{\citenamefont {Savinainen}\ \emph {et~al.}(2013)\citenamefont
  {Savinainen}, \citenamefont {Mäkynen}, \citenamefont {Nieminen},\ and\
  \citenamefont {Viiri}}]{savinainen_does_2013}%
  \BibitemOpen
  \bibfield  {author} {\bibinfo {author} {\bibfnamefont {A.}~\bibnamefont
  {Savinainen}}, \bibinfo {author} {\bibfnamefont {A.}~\bibnamefont
  {Mäkynen}}, \bibinfo {author} {\bibfnamefont {P.}~\bibnamefont {Nieminen}},\
  and\ \bibinfo {author} {\bibfnamefont {J.}~\bibnamefont {Viiri}},\ }\bibfield
   {title} {\bibinfo {title} {Does using a visual-representation tool foster
  students’ ability to identify forces and construct free-body diagrams?},\
  }\bibfield  {journal} {\bibinfo  {journal} {Physical Review Special Topics -
  Physics Education Research}\ }\textbf {\bibinfo {volume} {9}},\ \href
  {https://doi.org/10.1103/PhysRevSTPER.9.010104}
  {10.1103/PhysRevSTPER.9.010104} (\bibinfo {year} {2013})\BibitemShut
  {NoStop}%
\bibitem [{\citenamefont {Dufresne}\ \emph {et~al.}(1997)\citenamefont
  {Dufresne}, \citenamefont {Gerace},\ and\ \citenamefont
  {Leonard}}]{dufresne_solving_1997}%
  \BibitemOpen
  \bibfield  {author} {\bibinfo {author} {\bibfnamefont {R.~J.}\ \bibnamefont
  {Dufresne}}, \bibinfo {author} {\bibfnamefont {W.~J.}\ \bibnamefont
  {Gerace}},\ and\ \bibinfo {author} {\bibfnamefont {W.~J.}\ \bibnamefont
  {Leonard}},\ }\bibfield  {title} {\bibinfo {title} {Solving physics problems
  with multiple representations},\ }\href {https://doi.org/10.1119/1.2344681}
  {\bibfield  {journal} {\bibinfo  {journal} {The Physics Teacher}\ }\textbf
  {\bibinfo {volume} {35}},\ \bibinfo {pages} {270} (\bibinfo {year}
  {1997})}\BibitemShut {NoStop}%
\bibitem [{\citenamefont {Nguyen}\ \emph {et~al.}(2010)\citenamefont {Nguyen},
  \citenamefont {Gire}, \citenamefont {Rebello}, \citenamefont {Singh},
  \citenamefont {Sabella},\ and\ \citenamefont
  {Rebello}}]{nguyen_facilitating_2010}%
  \BibitemOpen
  \bibfield  {author} {\bibinfo {author} {\bibfnamefont {D.-H.}\ \bibnamefont
  {Nguyen}}, \bibinfo {author} {\bibfnamefont {E.}~\bibnamefont {Gire}},
  \bibinfo {author} {\bibfnamefont {N.~S.}\ \bibnamefont {Rebello}}, \bibinfo
  {author} {\bibfnamefont {C.}~\bibnamefont {Singh}}, \bibinfo {author}
  {\bibfnamefont {M.}~\bibnamefont {Sabella}},\ and\ \bibinfo {author}
  {\bibfnamefont {S.}~\bibnamefont {Rebello}},\ }\bibfield  {title} {\bibinfo
  {title} {Facilitating {Students}’ {Problem} {Solving} across {Multiple}
  {Representations} in {Introductory} {Mechanics}},\ }in\ \href
  {https://doi.org/10.1063/1.3515244} {\emph {\bibinfo {booktitle} {2010
  {Physics} {Education} {Research} {Conference} {Proceedings}}}}\ (\bibinfo
  {year} {2010})\ pp.\ \bibinfo {pages} {45--48}\BibitemShut {NoStop}%
\bibitem [{\citenamefont {Kohl}\ and\ \citenamefont
  {Finkelstein}(2005)}]{kohl_student_2005}%
  \BibitemOpen
  \bibfield  {author} {\bibinfo {author} {\bibfnamefont {P.~B.}\ \bibnamefont
  {Kohl}}\ and\ \bibinfo {author} {\bibfnamefont {N.~D.}\ \bibnamefont
  {Finkelstein}},\ }\bibfield  {title} {\bibinfo {title} {Student
  representational competence and self-assessment when solving physics
  problems},\ }\href {https://doi.org/10.1103/PhysRevSTPER.1.010104} {\bibfield
   {journal} {\bibinfo  {journal} {Physical Review Special Topics - Physics
  Education Research}\ }\textbf {\bibinfo {volume} {1}},\ \bibinfo {pages}
  {010104} (\bibinfo {year} {2005})}\BibitemShut {NoStop}%
\bibitem [{\citenamefont {McPadden}\ and\ \citenamefont
  {Brewe}(2015)}]{mcpadden_network_2015}%
  \BibitemOpen
  \bibfield  {author} {\bibinfo {author} {\bibfnamefont {D.}~\bibnamefont
  {McPadden}}\ and\ \bibinfo {author} {\bibfnamefont {E.}~\bibnamefont
  {Brewe}},\ }\bibfield  {title} {\bibinfo {title} {Network {Analysis} of
  {Students}' {Representation} {Use} in {Problem} {Solving}},\ }in\ \href
  {https://www.compadre.org/portal/items/detail.cfm?ID=13875} {\emph {\bibinfo
  {booktitle} {2015 {Physics} {Education} {Research} {Conference}
  {Proceedings}}}}\ (\bibinfo {year} {2015})\ pp.\ \bibinfo {pages}
  {219--222}\BibitemShut {NoStop}%
\bibitem [{\citenamefont {Gire}\ and\ \citenamefont
  {Price}(2015)}]{gire_structural_2015}%
  \BibitemOpen
  \bibfield  {author} {\bibinfo {author} {\bibfnamefont {E.}~\bibnamefont
  {Gire}}\ and\ \bibinfo {author} {\bibfnamefont {E.}~\bibnamefont {Price}},\
  }\bibfield  {title} {\bibinfo {title} {Structural features of algebraic
  quantum notations},\ }\bibfield  {journal} {\bibinfo  {journal} {Physical
  Review Special Topics - Physics Education Research}\ }\textbf {\bibinfo
  {volume} {11}},\ \href {https://doi.org/10.1103/PhysRevSTPER.11.020109}
  {10.1103/PhysRevSTPER.11.020109} (\bibinfo {year} {2015})\BibitemShut
  {NoStop}%
\bibitem [{\citenamefont {Bajracharya}\ \emph {et~al.}(2019)\citenamefont
  {Bajracharya}, \citenamefont {Emigh},\ and\ \citenamefont
  {Manogue}}]{bajracharya_students_2019}%
  \BibitemOpen
  \bibfield  {author} {\bibinfo {author} {\bibfnamefont {R.~R.}\ \bibnamefont
  {Bajracharya}}, \bibinfo {author} {\bibfnamefont {P.~J.}\ \bibnamefont
  {Emigh}},\ and\ \bibinfo {author} {\bibfnamefont {C.~A.}\ \bibnamefont
  {Manogue}},\ }\bibfield  {title} {\bibinfo {title} {Students' strategies for
  solving a multirepresentational partial derivative problem in
  thermodynamics},\ }\href
  {https://doi.org/10.1103/PhysRevPhysEducRes.15.020124} {\bibfield  {journal}
  {\bibinfo  {journal} {Physical Review Physics Education Research}\ }\textbf
  {\bibinfo {volume} {15}},\ \bibinfo {pages} {020124} (\bibinfo {year}
  {2019})},\ \bibinfo {note} {publisher: American Physical Society}\BibitemShut
  {NoStop}%
\bibitem [{\citenamefont {Hand}\ and\ \citenamefont
  {Choi}(2010)}]{hand_examining_2010}%
  \BibitemOpen
  \bibfield  {author} {\bibinfo {author} {\bibfnamefont {B.}~\bibnamefont
  {Hand}}\ and\ \bibinfo {author} {\bibfnamefont {A.}~\bibnamefont {Choi}},\
  }\bibfield  {title} {\bibinfo {title} {Examining the {Impact} of {Student}
  {Use} of {Multiple} {Modal} {Representations} in {Constructing} {Arguments}
  in {Organic} {Chemistry} {Laboratory} {Classes}},\ }\href
  {https://doi.org/10.1007/s11165-009-9155-8} {\bibfield  {journal} {\bibinfo
  {journal} {Research in Science Education}\ }\textbf {\bibinfo {volume}
  {40}},\ \bibinfo {pages} {29} (\bibinfo {year} {2010})}\BibitemShut {NoStop}%
\bibitem [{\citenamefont {Susac}\ \emph {et~al.}(2019)\citenamefont {Susac},
  \citenamefont {Bubic}, \citenamefont {Planinic}, \citenamefont {Movre},\ and\
  \citenamefont {Palmovic}}]{susac_role_2019}%
  \BibitemOpen
  \bibfield  {author} {\bibinfo {author} {\bibfnamefont {A.}~\bibnamefont
  {Susac}}, \bibinfo {author} {\bibfnamefont {A.}~\bibnamefont {Bubic}},
  \bibinfo {author} {\bibfnamefont {M.}~\bibnamefont {Planinic}}, \bibinfo
  {author} {\bibfnamefont {M.}~\bibnamefont {Movre}},\ and\ \bibinfo {author}
  {\bibfnamefont {M.}~\bibnamefont {Palmovic}},\ }\bibfield  {title} {\bibinfo
  {title} {Role of diagrams in problem solving: {An} evaluation of eye-tracking
  parameters as a measure of visual attention},\ }\href
  {https://doi.org/10.1103/PhysRevPhysEducRes.15.013101} {\bibfield  {journal}
  {\bibinfo  {journal} {Physical Review Physics Education Research}\ }\textbf
  {\bibinfo {volume} {15}},\ \bibinfo {pages} {013101} (\bibinfo {year}
  {2019})},\ \bibinfo {note} {publisher: American Physical Society}\BibitemShut
  {NoStop}%
\bibitem [{\citenamefont {Gire}\ \emph {et~al.}(2017)\citenamefont {Gire},
  \citenamefont {Wangberg},\ and\ \citenamefont
  {Wangberg}}]{gire_multiple_2017}%
  \BibitemOpen
  \bibfield  {author} {\bibinfo {author} {\bibfnamefont {E.}~\bibnamefont
  {Gire}}, \bibinfo {author} {\bibfnamefont {A.}~\bibnamefont {Wangberg}},\
  and\ \bibinfo {author} {\bibfnamefont {R.}~\bibnamefont {Wangberg}},\
  }\bibfield  {title} {\bibinfo {title} {Multiple tools for visualizing
  equipotential surfaces: {Optimizing} for instructional goals},\ }in\
  \href@noop {} {\emph {\bibinfo {booktitle} {Physics {Education} {Research}
  {Conference}}}}\ (\bibinfo {year} {2017})\BibitemShut {NoStop}%
\bibitem [{\citenamefont {Vignal}\ and\ \citenamefont
  {Wilcox}(2020)}]{vignal_comparing_2020}%
  \BibitemOpen
  \bibfield  {author} {\bibinfo {author} {\bibfnamefont {M.}~\bibnamefont
  {Vignal}}\ and\ \bibinfo {author} {\bibfnamefont {B.~R.}\ \bibnamefont
  {Wilcox}},\ }\bibfield  {title} {\bibinfo {title} {Comparing {Unprompted} and
  {Prompted} {Student}-{Generated} {Diagrams}}\ }(\bibinfo {year} {2020})\ pp.\
  \bibinfo {pages} {551--556},\ \bibinfo {note} {iSSN: 2377-2379}\BibitemShut
  {NoStop}%
\bibitem [{\citenamefont {Gupta}\ and\ \citenamefont
  {Elby}(2011)}]{gupta_beyond_2011}%
  \BibitemOpen
  \bibfield  {author} {\bibinfo {author} {\bibfnamefont {A.}~\bibnamefont
  {Gupta}}\ and\ \bibinfo {author} {\bibfnamefont {A.}~\bibnamefont {Elby}},\
  }\bibfield  {title} {\bibinfo {title} {Beyond {Epistemological} {Deficits}:
  {Dynamic} explanations of engineering students’ difficulties with
  mathematical sense-making},\ }\href
  {https://doi.org/10.1080/09500693.2010.551551} {\bibfield  {journal}
  {\bibinfo  {journal} {International Journal of Science Education}\ }\textbf
  {\bibinfo {volume} {33}},\ \bibinfo {pages} {2463} (\bibinfo {year}
  {2011})}\BibitemShut {NoStop}%
\bibitem [{\citenamefont {Wilcox}\ and\ \citenamefont
  {Pollock}(2015)}]{wilcox_student_2015}%
  \BibitemOpen
  \bibfield  {author} {\bibinfo {author} {\bibfnamefont {B.~R.}\ \bibnamefont
  {Wilcox}}\ and\ \bibinfo {author} {\bibfnamefont {S.~J.}\ \bibnamefont
  {Pollock}},\ }\bibfield  {title} {\bibinfo {title} {Student {Difficulties}
  with the {Dirac} {Delta} {Function}},\ }in\ \href
  {https://www.compadre.org/portal/items/detail.cfm?ID=13504} {\emph {\bibinfo
  {booktitle} {2014 {Physics} {Education} {Research} {Conference}
  {Proceedings}}}}\ (\bibinfo {year} {2015})\ pp.\ \bibinfo {pages}
  {271--274},\ \bibinfo {note} {iSSN: 2377-2379}\BibitemShut {NoStop}%
\bibitem [{\citenamefont {Knight}\ \emph {et~al.}(2015)\citenamefont {Knight},
  \citenamefont {Jones},\ and\ \citenamefont {Field}}]{knight_college_2015}%
  \BibitemOpen
  \bibfield  {author} {\bibinfo {author} {\bibfnamefont {R.~D.}\ \bibnamefont
  {Knight}}, \bibinfo {author} {\bibfnamefont {B.}~\bibnamefont {Jones}},\ and\
  \bibinfo {author} {\bibfnamefont {S.}~\bibnamefont {Field}},\ }\href@noop {}
  {\emph {\bibinfo {title} {college physics}}}\ (\bibinfo  {publisher} {Pearson
  Education UK},\ \bibinfo {year} {2015})\BibitemShut {NoStop}%
\bibitem [{\citenamefont {Giancoli}(2008)}]{giancoli_physics_2008}%
  \BibitemOpen
  \bibfield  {author} {\bibinfo {author} {\bibfnamefont {D.~C.}\ \bibnamefont
  {Giancoli}},\ }\href@noop {} {\emph {\bibinfo {title} {Physics for scientists
  and engineers with modern physics}}}\ (\bibinfo  {publisher} {Pearson
  Education},\ \bibinfo {year} {2008})\BibitemShut {NoStop}%
\bibitem [{\citenamefont {Hammer}\ \emph {et~al.}(2005)\citenamefont {Hammer},
  \citenamefont {Elby}, \citenamefont {Scherr},\ and\ \citenamefont
  {Redish}}]{hammer_resources_2005}%
  \BibitemOpen
  \bibfield  {author} {\bibinfo {author} {\bibfnamefont {D.}~\bibnamefont
  {Hammer}}, \bibinfo {author} {\bibfnamefont {A.}~\bibnamefont {Elby}},
  \bibinfo {author} {\bibfnamefont {R.~E.}\ \bibnamefont {Scherr}},\ and\
  \bibinfo {author} {\bibfnamefont {E.~F.}\ \bibnamefont {Redish}},\ }\bibfield
   {title} {\bibinfo {title} {Resources, framing, and transfer},\ }\href@noop
  {} {\bibfield  {journal} {\bibinfo  {journal} {Transfer of learning from a
  modern multidisciplinary perspective}\ }\textbf {\bibinfo {volume} {89}}
  (\bibinfo {year} {2005})}\BibitemShut {NoStop}%
\bibitem [{\citenamefont {Maries}\ \emph {et~al.}(2016)\citenamefont {Maries},
  \citenamefont {Lin},\ and\ \citenamefont {Singh}}]{maries_impact_2016}%
  \BibitemOpen
  \bibfield  {author} {\bibinfo {author} {\bibfnamefont {A.}~\bibnamefont
  {Maries}}, \bibinfo {author} {\bibfnamefont {S.-Y.}\ \bibnamefont {Lin}},\
  and\ \bibinfo {author} {\bibfnamefont {C.}~\bibnamefont {Singh}},\ }\bibfield
   {title} {\bibinfo {title} {The impact of students' epistemological framing
  on a task requiring representational consistency},\ }in\ \href
  {https://www.compadre.org/portal/items/detail.cfm?ID=14231} {\emph {\bibinfo
  {booktitle} {2016 {Physics} {Education} {Research} {Conference}
  {Proceedings}}}}\ (\bibinfo {year} {2016})\ pp.\ \bibinfo {pages}
  {212--215}\BibitemShut {NoStop}%
\bibitem [{\citenamefont {{Roy D. Pea}}(1997)}]{salomon_practices_1997}%
  \BibitemOpen
  \bibfield  {author} {\bibinfo {author} {\bibnamefont {{Roy D. Pea}}},\
  }\bibfield  {title} {\bibinfo {title} {Practices of {Distributed}
  {Intelligence} and {Designs} for {Education}},\ }in\ \href@noop {} {\emph
  {\bibinfo {booktitle} {Distributed cognitions: psychological and educational
  considerations}}},\ \bibinfo {series and number} {Learning in doing},\
  \bibinfo {editor} {edited by\ \bibinfo {editor} {\bibfnamefont
  {G.}~\bibnamefont {Salomon}}}\ (\bibinfo  {publisher} {Cambridge University
  Press},\ \bibinfo {address} {Cambridge [Cambridgeshire] ; New York},\
  \bibinfo {year} {1997})\ \bibinfo {edition} {1st}\ ed.,\ pp.\ \bibinfo
  {pages} {47--87}\BibitemShut {NoStop}%
\bibitem [{\citenamefont {Otero}(2002)}]{otero_conceptual_2002}%
  \BibitemOpen
  \bibfield  {author} {\bibinfo {author} {\bibfnamefont {V.~K.}\ \bibnamefont
  {Otero}},\ }\bibfield  {title} {\bibinfo {title} {Conceptual {Development}
  and {Context}: {How} {Do} {They} {Relate}?},\ }in\ \href
  {https://www.compadre.org/portal/items/detail.cfm?ID=11218} {\emph {\bibinfo
  {booktitle} {2002 {Physics} {Education} {Research} {Conference}
  {Proceedings}}}}\ (\bibinfo {year} {2002})\BibitemShut {NoStop}%
\bibitem [{\citenamefont {Stemler}(2015)}]{stemler_content_2015}%
  \BibitemOpen
  \bibfield  {author} {\bibinfo {author} {\bibfnamefont {S.~E.}\ \bibnamefont
  {Stemler}},\ }\bibfield  {title} {\bibinfo {title} {Content analysis},\
  }\href@noop {} {\bibfield  {journal} {\bibinfo  {journal} {Emerging Trends in
  the Social and Behavioral Sciences: An Interdisciplinary, Searchable, and
  Linkable Resource}\ ,\ \bibinfo {pages} {1}} (\bibinfo {year}
  {2015})}\BibitemShut {NoStop}%
\bibitem [{\citenamefont {Cresswell}(2014)}]{cresswell_research_2014}%
  \BibitemOpen
  \bibfield  {author} {\bibinfo {author} {\bibfnamefont {J.~W.}\ \bibnamefont
  {Cresswell}},\ }\bibfield  {title} {\bibinfo {title} {Research {Design}:
  {Qualitative}, {Quantitative}, and {Mixed} {Methods} {Approaches}}\
  }(\bibinfo  {publisher} {SAGE Publications, Inc.},\ \bibinfo {year} {2014})\
  \bibinfo {edition} {4th}\ ed.,\ pp.\ \bibinfo {pages} {183--213}\BibitemShut
  {NoStop}%
\end{thebibliography}%

\end{document}